\documentclass[10pt, a4paper, onecolumn]{article}
\usepackage{amsmath,amssymb,amsfonts,cite}
\usepackage{pdflscape}
\usepackage{graphicx}
\usepackage{pict2e}
\usepackage{enumerate}
\usepackage{setspace}
\usepackage{xcolor}
\usepackage{multirow}
\usepackage{amsbsy} 
\usepackage{bm}
\usepackage{textcomp}
\usepackage{gensymb}
\usepackage{boxedminipage}
\usepackage{framed}
\usepackage{caption}
\usepackage[affil-it]{authblk} 
\usepackage{etoolbox}
\usepackage{lmodern}
\usepackage{booktabs}
\usepackage{tabularx}
\usepackage{siunitx}
\usepackage{array}
\usepackage{authblk}
\usepackage{indentfirst}
\usepackage{hyperref}
\usepackage{color}
\usepackage{float}
\usepackage[title]{appendix}
\date{}

\begin{document}
\title{Segmentation based tracking of cells in 2D+time microscopy images of macrophages}
\author[1]{Seol Ah Park}
\author[2]{Tamara Sipka}
\author[1]{Zuzana Kriva}
\author[2]{George Lutfalla}
\author[2]{Mai Nguyen-Chi}
\author[1]{Karol Mikula}
\affil[1]{\footnotesize Department of Mathematics and Descriptive Geometry, Slovak University of Technology in Bratislava, Slovakia}
\affil[2]{\footnotesize DIMNP, CNRS, Univ. Montpellier, Montpellier, France}
\maketitle
\begin{abstract}
The automated segmentation and tracking of macrophages during their migration are challenging tasks due to their dynamically changing shapes and motions. This paper proposes a new algorithm to achieve automatic cell tracking in time-lapse microscopy macrophage data. 
First, we design a segmentation method employing space-time filtering, local Otsu's thresholding, and the SUBSURF (subjective surface segmentation) method. Next, the partial trajectories for cells overlapping in the temporal direction are extracted in the segmented images. Finally, the extracted trajectories are linked by considering their direction of movement.
The segmented images and the obtained trajectories from the proposed method are compared with those of the semi-automatic segmentation and manual tracking.
The proposed tracking achieved  97.4\% of accuracy for macrophage data under challenging situations, feeble fluorescent intensity, irregular shapes, and motion of macrophages.
We expect that the automatically extracted trajectories of macrophages can provide pieces of evidence of how macrophages migrate depending on their polarization modes in the situation, such as during wound healing.

\end{abstract}

\section{Introduction}
Since the 17th century and the first microscopes, biologists have dedicated enormous efforts to understanding cellular behaviors within living animals \cite{needham1745}. Embryologists have first described how cellular movements shape embryonic development, but immunologists soon realized that by using microscopy, they could have access to the behavior of specialized, highly mobile cells that play crucial roles in immunity \cite{metchnikoff1892}. With the recent development of video microscopy, the diversification of confocal microscopy techniques, and the constant improvement of the sensitivity, resolution, and speed of acquisitions of microscopes \cite{huang2021frontier,cuny2022live}, biologists are now generating huge sets of data that need automated processing to extract significant data to describe the integrated process and understand underlying rules. Thanks to the contributions of theoreticians and modelers, biologists can now integrate these imaging data with biochemical and genetic data to propose integrated models of cellular behaviors and even to offer integrated models of the development of as complex organisms as vertebrates \cite{delile2017,dutta2021}.
\newline
\indent
Identifying (segmenting) and tracking individual cells is challenging because cells divide, move, and change their shapes
during their journey in the developing embryo. Many efforts have been dedicated to developing software to track cells during embryonic development, and robust solutions are now available \cite{emami2021}. Some of these solutions are compatible with the study of other situations where cells are either moving in an organism (the heart) or in a moving organism (neurons in foraging worms \cite{wen2021}), but some specific cellular populations, due to their particular behaviors, are challenging to identify and track during their journey within a living animal. This is the case of macrophages, one of the fastest-moving cellular populations with more irregular shapes and movements.
\newline
\indent
Macrophages have protective roles in immune defense, homeostasis, and tissue repair, but they also contribute to the progression of many pathologies like cancers, inflammatory diseases, and infections \cite{wynn2013}. The key feature of macrophages is their remarkable dynamic plasticity. They respond to changing environments by constantly adopting specific phenotypes and functions defined as M1 and M2, which are the two extremes of a continuum of polarization states \cite{martinez2014}. In the early stage of inflammation, M1 macrophages have been shown to accumulate at the wound/infection site where they initiate a pro-inflammatory response showing highly phagocytic and removing any pathogens or debris \cite{benoit2008,daley2010,koh2011,hesketh2017}. During the resolution of inflammation, they switch to M2 macrophages which mediate anti-inflammatory response and participate in tissue remodeling and repair \cite{stables2011,klinkert2017,shook2016,hesketh2017}.
Some studies have reported that the different functions of M1/M2 macrophages seem to be related to shapes and migration \cite{friedl2008,van2010,barros2017,cui2018}. M1 macrophages are more rounded and flat shapes than M2 macrophages, showed by elongated shapes \cite{cui2018,van2010}. In addition to the variable morphology, macrophages are known to have two migration modes: amoeboid and mesenchymal. Amoeboid migration has a fast speed in a largely adhesion-independent manner, mainly observed for M1 macrophages. In contrast, mesenchymal migration is slower and more directional in the presence of strong adhesion mainly observed for M2 macrophages \cite{friedl2008,van2010,barros2017}. So far, the relationship between the macrophage activation and migration modes involving the change of macrophages' shapes {\textit{in vivo}} is still unclear. Image segmentation and cell tracking in macrophage data can be the first steps to analyzing the characteristics of macrophages \cite{holmes2012,kadirkamanathan2012}.
\newline
\subsection*{Related works and contribution to macrophage segmentation}
\indent 
Segmentation of macrophages has been previously studied performing a filter-based method \cite{wagner2019}, image-based machine learning \cite{rostam2017}, anglegram analysis \cite{solis2018}, etc. 
Also, deep learning-based segmentation methods have been developed for various types of cells \cite{ronneberger2015,schmidt2018,falk2019,lugagne2020,stringer2021,mandal2021}. 
U-Net\cite{ronneberger2015}, Cellpose\cite{stringer2021}, and Splinedist\cite{mandal2021} are designed for segmentation of general shapes of cells in microscopy data and have shown a high performance. However, it is still challenging to segment macrophages due to their varying nature, extreme irregularity of shapes, and variability of image intensity inside macrophages.
In \cite{park2020}, we have proposed a macrophage segmentation method that combines thresholding methods with the SUBSURF approach requiring no cell nuclei center or other reference information. However, a problem occurs when attempting to segment macrophages in time-lapse data since the segmentation parameters are not always suitable for macrophages in all time frames. In this paper, first, we improve the ability to detect macrophages with low image intensity by applying space-time filtering, which considers the temporal coherence of time-lapse data \cite{sarti1999}. Second, Otsu’s method is implemented in local windows to deal with cases where each macrophage has a substantially different image intensity range. Similarly, as in \cite{park2020}, the SUBSURF method \cite{sarti2000} is applied to eliminate the remaining noise and to smoothen the boundaries of the macrophages resulting from space-time filtering and the thresholding method (Fig. \ref{fig_outline}).   
\subsection*{Related works in cell tracking}
\indent Automatic cell tracking in microscopy images has been investigated and various methods \cite{jaqaman2008,kadirkamanathan2012, amat2014, mikula2015, faure2016, spir2016,tinevez2017,lugagne2020,aragaki2022} have been proposed. 
The tracking algorithm using linear assignment problem (LAP) \cite{jaqaman2008,tinevez2017} is computationally efficient and has shown good performance, especially for Brownian motion. However, it can be less accurate if many cells are densely distributed or if some cells suddenly move toward the other nearby cells.
The studies \cite{mikula2015, spir2016} performed cell tracking during zebrafish embryogenesis by finding a centered path in the spatio-temporal segmented structure.
In \cite{faure2016}, a workflow was designed, from the image acquisition to cell tracking, and applied to 3D+time microscopy data of the zebrafish embryos.
Those methods show outstanding performance in the case of embryogenesis.
The keyhole tracking algorithm that anticipates the most probable position of a cell at the next time slice has been proposed and applied to red blood cells, neutrophils, and macrophages \cite{reyes2008,reyes2009,henry2013}. 
Moreover, deep learning-based motion tracking in microscopy images has been studied for various types of biological objects with different learning approaches \cite{cooper2017nuclitrack,ulman2017objective,ramesh2018semi,hernandez2018cell,tsai2019usiigaci, arts2019particle,wang2020deep,iriya2020rapid,masoudi2020instance,liu2021survey}.
For instance, the method in \cite{ramesh2018semi} trains the networks by utilizing semi-supervised learning to predict cell division. Usiigaci \cite{tsai2019usiigaci} segments individual cells providing each unique ID to them with a Mask R-CNN model, then the method links the cells by given IDs. The methods by training image sequences using  LSTM (long short-term memory) networks have shown their performance for tracing nuclear proteins \cite{arts2019particle} and bacteria \cite{iriya2020rapid}.
In \cite{wang2020deep}, the algorithm to solve linear assignment problems in tracking is trained with a deep reinforcement learning (DRL)-based method.
\subsection*{Contribution to macrophage tracking and outline}
\begin{figure}[ht]
	\centering
   \includegraphics[scale=0.55]{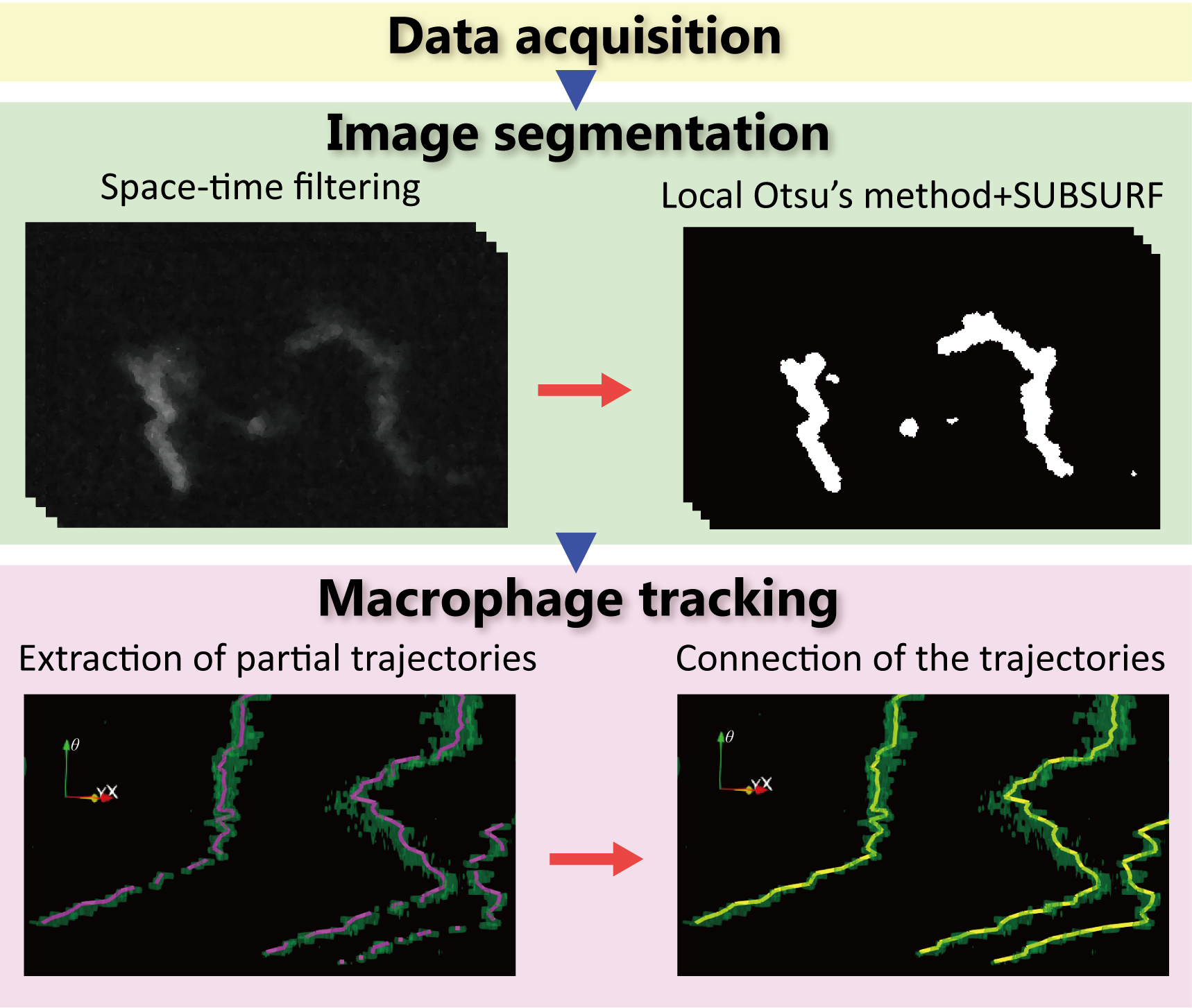}
    \caption{Procedure of macrophage tracking in 2D+time data.}
   \label{fig_outline}
\end{figure}

Although various cell tracking methods have been studied, there is still a need for more accurate tracking of erratic movements, such as macrophages. The cell tracking studied in this paper deals with macrophages which undergo fast and complicated motion. It results in non-overlapping cells in the time direction, and in many cases, one can observe a ``random movement''.
This paper proposes a tracking method that covers the situations of a large number of macrophages and their complex motion. The first step is to extract the cell trajectories from their shapes overlapping in time. By this approach, we often obtain only partial trajectories because not always a segmented macrophage overlaps with its corresponding cell in the next/previous frame of the video. Next, we connect endpoints of partial trajectories corresponding to macrophages that do not overlap in time. For this, the tangent calculation is used to estimate the direction of macrophages at the endpoints of the partial trajectories. Fig. \ref{fig_outline} illustrates briefly all steps of the proposed method yielding macrophage tracking.
\newline
\indent 
The mathematical descriptions of the proposed method are illustrated in Materials and methods. The performances of the macrophage segmentation and tracking are shown in Results.
In this section, the proposed segmentation method provides the approximate shapes of macrophages, indicating that it can reasonably be the first tracking step. Also, proposed tracking shows that connecting the centers of macrophages by considering the direction of movement works properly for tracing fast-moving macrophages. The tracking performance is validated visually and quantitatively, showing how obtained trajectories are close to the manually extracted trajectories. In Discussion, we summarize the results of the proposed method and discuss limitations, future works, and possible applications.

\renewcommand{\figurename}{Fig.}

\section{Materials and methods}
\subsection{Image acquisition and preparation}
\indent 
The proposed method is applied to two representative datasets. In both datasets, a three days old transgenic zebrafish larva (Tg(mpeg1:Gal4/-UAS:Kaede)) is used and imaged using a spinning disk confocal microscope. The green fluorescent protein Kaede is indirectly expressed under the control of macrophage-specific promoter mpeg1 so that macrophages produce the green fluorescent protein in their cytoplasm. In the first dataset, migrating macrophages are imaged from $1$ hour to $6$ hours post-amputation $(1-6$ $hpA)$ for the caudal fin fold with a time step of $4$ minutes and a $z$ step of $4$ $\mu m$. 
In the second dataset, macrophages are imaged from $30$ minutes post-amputation to $6$ hours post-amputation $(0.5-6$ $hpA)$ with the imaging time step of $2$ minutes and $z$ step of $1$ $\mu m$. The pixel size is $0.326$ $\mu m$ and $0.347$ $\mu m$ in the first and second datasets, respectively. For the numerical experiments, we used the 2D+time projection images, where the three-dimensional (3D) microscopy images are projected onto a plane with the maximum intensity of the 3D dataset in every pixel selected. Due to the image acquisition speed in the second dataset, the exposition time and fluorescence intensity are reduced, resulting in a low signal-to-noise ratio.
\newline
\indent 
We perform the histogram crop from the acquired images to ignore the noise effects from very high image intensity in a small pixel area. 
In the case of this noise, the number of pixels is very small compared with the image size. Therefore, in the histogram, a tiny peak positioned at the highest image intensity corresponds to this type of noise.
To ignore it, this tiny peak will be cropped in the histogram.  
The steps of the histogram crop are the following.
\begin{enumerate}
    \item The first step is the estimation of the noise size relative to the image size. 
    Let us consider the noise accounts for $p_{\text{noise}}$ percent of the total number of pixels $N_{\text{tot}}$.
    Then, the number of pixels for the noise $N_{\text{noise}}$ equals to $N_{\text{tot}}\times p_{\text{noise}}$.
    \item In the histogram, the number of pixels from the maximum intensity in descending order is counted because we want to remove the small noise having the highest image intensity. 
    Let us denote the counted number of the pixels as $N_{\text{des}}(I)$. 
    For example, $N_{\text{des}}(I)$ for the maximum image intensity $I_{\text{max}}$ equals to the number of pixels of the image intensity $I_{\text{max}}$.
    Likewise, for the minimum image intensity $I_{\text{min}}$, the counted number of pixels from $I_{\text{max}}$ to $I_{\text{min}}$ is $N_{\text{des}}(I_{\text{min}})=N_{\text{tot}}$. 
    \item Finally, when $N_{\text{des}}(I^{*})=N_{\text{noise}}$ is satisfied, counting is stopped. The new maximum intensity $I_{\text{new;max}}$ is set by searching for the maximum intensity smaller than $I^{*}$.  The image intensity ranging from $I^{*}$ to $I_{\text{max}}$ is changed to $I_{\text{new;max}}$.
\end{enumerate}
In the supplementary material (\url{https://doi.org/10.1016/j.compbiomed.2022.106499}), an example of histograms in the presence of the spot noise and after the histogram crop is depicted.
We apply the histogram crop only to the second dataset as $p_{\text{noise}}=0.001$.
\newline
\indent
After the histogram crop, the image intensity is scaled to the interval $[0,1]$ for applying space-time filtering. Then, the images obtained from space-time filtering are rescaled to the interval $[0,255]$ to simply perform the local Otsu's method since histograms of images are usually described by the discrete distribution in a finite interval. 
To apply the SUBSURF method, two types of images are used; one is the original images after the histogram crop with the interval $[0,1]$, and the other is the output of the local Otsu's method.

\subsection{Segmentation of macrophages in microscopy videos}
\subsubsection{Space-time filtering}
In the datasets processed by the methods presented in this paper, the macrophages do not always have similar image intensities.
Some can hardly be recognized due to their weak image intensity in static images, but they can be recognized in videos, as human eyes consider temporal information to distinguish objects. However, the traditional segmentation method dealing with static images does not view temporal information. Therefore, it is difficult to detect and segment macrophages if the image intensity of macrophages is similar to the background.
The filtering method that can utilize temporal coherence was introduced in \cite{sarti1999}, where the regularized Perona--Malik model and scalar function $clt$, measuring the coherence of objects in time slices, are combined. 
The term $clt$ means a ``curvature of Lambertian trajectory'' \cite{alvarez1993,alvarez1994,guichard1994} and vanishes for points (of an object) that preserve the intensity and move on a smooth trajectory in a temporal direction.
\newline 
\indent 
In the following, let a sequence of time slices be given on the interval [$0,\theta_{F}$], and $\theta$ will denote a particular time slice.
The PDE representing nonlinear diffusion filtering is written as
\begin{equation} \label{sp_filter}
    \frac{\partial u}{\partial t} = clt(u)\nabla \cdot \bigl(g(|\nabla G_{\sigma}*u|)\nabla u\bigr),
\end{equation}
where $t$ denotes the scale, the amount of filtering and $u(t,x_1,x_2,\theta)$ is the unknown real function which is defined on $[0,T_{F}] \times \Omega \times  [0,\theta_{F}] $, $\textbf{x}=(x_1,x_2) \in \Omega \subset R^{2}$. In $|\nabla G_{\sigma}*u|$, the ``$*$'' stands for the convolution operator.

\noindent The initial condition given by  
\begin{equation} \label{initial_filter}
    u(0,\textbf{x},\theta)=u^{0}(\textbf{x},\theta),
\end{equation}
represents the processed 2D+time video.
The $clt(u)$ function is defined as in \cite{alvarez1993, sarti1999} by formula
\begin{equation} \label{clt}
clt(u)=\min_{\boldsymbol{w_1}, \boldsymbol{w_2}}\frac{1}{(\Delta \theta)^2} \bigl(|<\nabla u,\boldsymbol{w_1}-\boldsymbol{w_2}>|+|u(\mathbf{x}-\boldsymbol{w_1}, \theta-\Delta \theta)-u(\mathbf{x},\theta)|+|u(\mathbf{x}+\boldsymbol{w_2}, \theta+\Delta \theta)-u(\mathbf{x},\theta)|\bigr),
\end{equation}
where $\boldsymbol w_1$, $\boldsymbol w_2$ are arbitrary vectors in 2D space, and $\Delta\theta$ is the time increment between discrete time slices. Here, $<\boldsymbol{a},\boldsymbol{b}>$ denotes Euclidean  scalar product of $\boldsymbol{a}$ and $\boldsymbol{b}$.
The function $g$ is the so-called edge detector function and is defined by
\begin{equation} \label{g_function}
    g(s)=\frac{1}{1+Ks^2}, \ K>0,
\end{equation}
where $K$ is a constant to control the sensitivity of $s$ \cite{perona1990scale}. 
Finally, $G_\sigma$ is a Gaussian function with variance $\sigma$, which is used for pre-smoothing by convolution.
\noindent Let us denote by $u_{k}^{n}$ a numerical solution in the $k^{th}$ frame of the image sequence in the $n^{th}$ discrete filtering (scale) step $n\tau_{F}$ with the step size $\tau_{F}$, i.e.,
\begin{equation}
    u^{n}_{k}(\mathbf{x})=u(n\tau_{F},\mathbf{x},k\Delta \theta).
\end{equation}
By using the semi-implicit scheme \cite{sarti1999}, Equation \ref{sp_filter} is discretized as follows

\begin{equation} \label{sp_filter_discretized}
    \frac{u_{k}^{n+1}-u_{k}^{n}}{\tau_{F}}=clt(u_{k}^{n})\nabla\cdot\bigl(g(|\nabla u_{k}^{\sigma;n}|)\nabla u_{k}^{n+1}\bigr),
\end{equation}
where $g(|\nabla u_{k}^{\sigma;n}|)=g(|\nabla G_{\sigma}*u_{k}^{n}|)$.
From Equation \ref{clt}, the discretization of $clt(u_{k}^{n})$ in the point $\textbf{x} \in \Omega$ can be written as

\begin{equation} \label{clt_dist}
    clt(u_{k}^{n})= \min_{\boldsymbol{w_1}, \boldsymbol{w_2}}\frac{1}{\Delta\theta^{2}} \bigl(|<\nabla u_{k}^{n}, \boldsymbol{w_1}-\boldsymbol{w_2}>|+
    |u_{k-1}^{n}(\mathbf{x}-\boldsymbol{w_1})-u_{k}^{n}(\mathbf{x})|+
    |u_{k+1}^{n}(\mathbf{x}+\boldsymbol{w_2})-u_{k}^{n}(\mathbf{x})|\bigr).  
\end{equation}
For space discretization, we use the finite volume method with finite volume (pixel) side $h$. Let us consider that a point $\textbf{x}$ is a center of a pixel $(i,j)$ and let us denote by $\mathcal{V}_{i,j}$ a finite volume corresponding to pixel $(i,j)$, $i= 1, \cdots M$, $j= 1, \cdots N $.
The quantity $clt(u_{k}^{n})$ is considered constant in finite volumes. Then, Equation \ref{sp_filter_discretized} is integrated with the finite volume $\mathcal{V}_{i,j}$, and by using Green's theorem, we get 
\begin{equation}
    \int_{\mathcal{V}_{i,j}}\frac{u_{k}^{n+1}-u_{k}^{n}}{\tau_{F}} d\mathbf{x}
    =clt(u_{k}^{n})\int_{\partial \mathcal{V}_{i,j}}g(|\nabla u_{k}^{\sigma;n}|)\mathbf{\nabla}u_{k}^{n+1}\cdot \mathbf{n}_{i,j}dS,
\end{equation}
where $\mathbf{n}_{i,j}$ is a unit outward normal vector to the boundary of $\mathcal{V}_{i,j}$.
The gradient of $u$ on the pixel edges can be approximated by computing the average values of neighboring pixels. By using the diamond cell approach \cite{mikula2009}, we compute the average of neighboring pixel values in the corners of the pixel $(i,j)$ as follows (see also Figure S2 in the supplement materials \url{ https://doi.org/10.1016/j.compbiomed.2022.106499}). 
\begin{equation} \label{average_gradient}
\begin{aligned}
u^{1,1}_{i,j,k}=\frac{1}{4}(u^{n}_{i,j,k}+u^{n}_{i,j+1,k}+u^{n}_{i+1,j,k}+u^{n}_{i+1,j+1,k}),\\
u^{1,-1}_{i,j,k}=\frac{1}{4}(u^{n}_{i,j,k}+u^{n}_{i+1,j,k}+u^{n}_{i,j-1,k}+u^{n}_{i+1,j-1,k}),\\
u^{-1,-1}_{i,j,k}=\frac{1}{4}(u^{n}_{i,j,k}+u^{n}_{i-1,j,k}+u^{n}_{i,j-1,k}+u^{n}_{i-1,j-1,k}), \\
u^{-1,1}_{i,j,k}=\frac{1}{4}(u^{n}_{i,j,k}+u^{n}_{i,j+1,k}+u^{n}_{i-1,j,k}+u^{n}_{i-1,j+1,k}). \\
\end{aligned}
\end{equation}
The gradient of $u^{n}_{i,j,k}$ in $n^{th}$ filtering step, for a pixel $(i,j)$ in $k^{th}$ frame of the image sequence, is computed at the center of edges of the pixel \cite{mikula2009},
\begin{equation} \label{gradient}
\begin{aligned}
\nabla^{1,0}u^{n}_{i,j,k}=\frac{1}{h}(u^{n}_{i+1,j,k}-u^{n}_{i,j,k},u^{1,1}_{i,j,k}-u^{1,-1}_{i,j,k}),\\
\nabla^{0,-1}u^{n}_{i,j,k}=\frac{1}{h}(u^{1,-1}_{i,j,k}-u^{-1,-1}_{i,j,k},u^{n}_{i,j-1,k}-u^{n}_{i,j,k}),\\
\nabla^{-1,0}u^{n}_{i,j,k}=\frac{1}{h}(u^{n}_{i-1,j,k}-u^{n}_{i,j,k},u^{-1,1}_{i,j,k}-u^{-1,-1}_{i,j,k}),\\
\nabla^{0,1}u^{n}_{i,j,k}=\frac{1}{h}(u^{1,1}_{i,j,k}-u^{-1,1}_{i,j,k},u^{n}_{i,j+1,k}-u^{n}_{i,j,k}),\\
\end{aligned}
 \end{equation}
where $h$ denotes the pixel size.
With the set of grid neighbors $N_{i,j}$ that consists of all $(l,m)$ of $\mathcal{V}_{i,j}$, such that $l,m \in \{-1,0,1\}$, $|l|+|m|=1$, the final discretized form of Equation \ref{sp_filter} is written as

\begin{equation} \label{final_clt}
    u_{i,j,k}^{n+1}=u_{i,j,k}^{n}+\frac{\tau_{F}}{h^2} \ clt(u_{i,j,k}^{n})\sum_{|l|+|m|=1}g(|\nabla^{l,m} u_{i,j,k}^{\sigma;n}|)(u^{n+1}_{i+l,j+m,k}-u^{n+1}_{i,j,k}).
\end{equation}
For solving Equation \ref{final_clt}, the successive over-relaxation (SOR) method is used.  The SOR method is an iterative method for solving a linear system of equations as a variant of the Gauss--Seidel method \cite{barrett1994templates}. In our simulations, the relaxation factor of the SOR method was set to $1.8$ and the calculation was stopped when $\sum_{i=1}^{M}\sum_{j=1}^{N}|u_{i,j,k}^{n+1}-u_{i,j,k}^{n}| < 0.001$ for every $k$.

\subsubsection{Local Otsu thresholding} \label{sec_localotsu}
It has been shown that the traditional Otsu thresholding technique, which selects a threshold globally (global Otsu's method), works well for some shapes of macrophages in \cite{park2020}. However, global Otsu's method does not work for all macrophages if there is a wide range of macrophage image intensity. When cells have a huge variability of shapes, sizes, and intensities, local thresholding techniques can be a powerful segmentation tool  \cite{korzynska2013}. 
We, therefore, apply Otsu's method in local windows to realize the benefits of both Otsu's method \cite{otsu1979} and local thresholding techniques \cite{korzynska2013, saddami2019}.
In the global Otsu's method, the two classes which represent objects and the background are firstly defined with the help of a general threshold value $T_{r}$. Then the optimal threshold is obtained by finding a particular threshold value $T_{r}^{*}$ that maximizes the between-class variance of the two classes. For local Otsu's method, we calculate the optimal threshold in a certain window of size $s \times s$ centered in $(i,j)$ for every pixel. 
In the local window  $W_{i,j}$, the gray-level histogram is normalized, and a probability distribution is regarded as
\begin{equation} \label{otsu_prob}
p_{r}=n_{r}/N, \quad \sum_{r=0}^{L}p_{r}=1,
 \end{equation}
where $n_{r}$ is the number of pixels of intensity $r$ in $W_{i,j}$, $N=s^{2}$ and $L$ is the maximum image intensity.
Then, the probabilities of background and foreground in $W_{i,j}$ are given by

\begin{equation} \label{localotsu_w}
\begin{aligned}
\omega_{0}(T_{i,j})=\sum^{T_{i,j}}_{r=0}p_{r}, \quad \omega_{1}(T_{i,j})&=\sum^{L}_{r=T_{i,j}+1}p_{r}=1-\omega_{0}(T_{i,j}),
\end{aligned}
 \end{equation}
and means of background and foreground are given by

\begin{equation} \label{localotsu_mu}
\begin{aligned}
\mu_{0}(T_{i,j})&=\frac{1}{\omega_{0}(T_{i,j})}\sum^{T_{i,j}}_{r=0}rp_{r}, \\
\mu_{1}(T_{i,j})&=\frac{1}{\omega_{1}(T_{i,j})}\sum^{L}_{r=T_{i,j}+1}rp_{r}=\frac{\mu_{\text{tot}}-\mu_{0}(T_{i,j})\omega_{0}(T_{i,j})}{1-\omega_{0}(T_{i,j})}, \\
\end{aligned}
 \end{equation}
where $\mu_{\text{tot}}=\sum^{L}_{r=0}rp_{r}$.
Finally, the between-class variance, the variance between classes of foreground and the background, related to the pixel $(i,j)$ is defined as \cite{otsu1979} 
\begin{equation} \label{localotsu_variance_ori}
\sigma_{B}^{2}(T_{i,j})=\omega_{0}(T_{i,j})(\mu_{0}(T_{i,j})-\mu_{\text{tot}})^{2}+\omega_{1}(T_{i,j})(\mu_{1}(T_{i,j})-\mu_{\text{tot}})^{2}
 \end{equation}
which simplifies to

\begin{equation} \label{localotsu_sigme}
\sigma_{B}^{2}(T_{i,j})=\frac{\bigl(\mu_{\text{tot}}\omega_{0}(T_{i,j})-\mu_{0}(T_{i,j})\omega_{0}(T_{i,j})\bigr)^{2}}{\omega_{0}(T_{i,j})\bigl(1-\omega_{0}(T_{i,j})\bigr)},
 \end{equation}
and the optimal threshold $T_{i,j}^*$ is given by

\begin{equation} \label{localotsu_variance}
    \sigma_{B}^{2}(T^{*}_{i,j})=\max_{0 \leq T_{i,j} < L} \sigma_{B}^{2}(T_{i,j}).
\end{equation}
At the boundary of the image, mirroring is applied.
In the case where the local window contains only the background, the histogram completely loses its bi-modality, with the threshold $T^{*}_{i,j}$ representing some noise level. To obtain a reasonable threshold, we determine whether the local window is only located in the background or not by considering the relative difference between the mean levels of the two classes representing the object and the background. Let us consider that $\mu_{0}(T^{*}_{i,j})$ and $\mu_{1}(T^{*}_{i,j})$ are the mean levels of the background and the object, respectively, based on the threshold $T^{*}_{i,j}$. 
If $|\mu_{0}(T^{*}_{i,j})-\mu_{1}(T^{*}_{i,j})| < \varepsilon$, $\varepsilon$ is very small, then the two classes cannot be properly separated, and it is reasonable to conclude that the local window is located in the background.
In other words, the local window $W_{i,j}$ is considered as including an object when the following condition is fulfilled:

\begin{equation} \label{local_otsu_eval}
    \frac{|\mu_{0}(T^{*}_{i,j})-\mu_{1}(T^{*}_{i,j})|}{\mu_{0}(T^{*}_{i,j})} > \delta,
\end{equation}
where the relative difference is considered since the background noise level is different in each time slice.
Here, $\delta$ is a parameter to check whether the local window $W_{i,j}$ contains macrophages or not. If there is a part of macrophages in $W_{i,j}$, the relative difference in Equation \ref{local_otsu_eval} will have a larger value than $\delta$.
Finally, the binarized images are obtained by defining
\begin{equation} \label{binarized}
  B(i,j)=
	\begin{cases}
	1,  I(i,j) > T^{*}_{i,j} \ \text{and Equation \ref{local_otsu_eval}} \enspace \text{is fulfilled}  \\
	0, \text{otherwise} \\ 
 \end{cases}  
\end{equation}
where $I(i,j)$ is the image intensity of the pixel $(i,j)$ and $ T^{*}_{i,j}$ is given by Equation \ref{localotsu_variance}.

\subsubsection{SUBSURF method}
The SUBSURF method can effectively complete missing parts of boundaries, 
join adjacent level lines, and rapidly remove noise \cite{sarti2000}. 
In particular, this method has previously been shown to be useful for segmenting macrophage data \cite{park2020}.
The SUBSURF method is described by

 \begin{equation} \label{subsurf}
\frac{\partial u}{\partial t}=|\mathbf{\nabla}u|\mathbf{\nabla}\cdot\biggl(g\frac{\mathbf{\nabla}u}{|\mathbf{\nabla}u|}\biggr),
 \end{equation}
where $u$ is a evolving level set function, $g=g(|\nabla G_{\sigma}*I^{0}|)$, and $s=|\nabla I^{0}_{\sigma}|$ in  Equation \ref{g_function}. Here, $I^{0}$ is the original image, and $I^{0}_{\sigma}$ is the pre-smoothed image with a Gaussian filter.
The SUBSURF is applied independently to the 2D images for every time frame. Therefore, we solve the unknown function $u(t,\textbf{x})$, where $(t,\textbf{x})$ $\in [0,T_{S}] \times \Omega, \textbf{x} \in \Omega \subset R^{2}$.
The time discretization of Equation \ref{subsurf} is given by the semi-implicit scheme
 \begin{equation} \label{t_subsurf}
\frac{u^{n+1}-u^{n}}{\tau_{S}}=|\mathbf{\nabla}u^{n}|_{\epsilon}\mathbf{\nabla}\cdot\biggl(g\frac{\mathbf{\nabla}u^{n+1}}{|\mathbf{\nabla}u^{n}|_{\epsilon}}\biggr),
 \end{equation}
where $\tau_{S}$ is the scale step. Here, $|\mathbf{\nabla}u^{n}|$ is regularized 
using the Evans--Sprucks approach \cite{evans1992} as $|\mathbf{\nabla}u^{n}|_{\epsilon} = \sqrt{|\mathbf{\nabla}u^{n}|^2 +\epsilon^{2}}$, 
where $\epsilon$ is a small arbitrary constant. 
The space is discretized by a finite volume square grid with the pixel size $h$.
For $\mathcal{V}_{i,j}$, Equation \ref{subsurf} is integrated and using Green's formula we get
 \begin{equation} \label{s_subsurf}
\int_{\mathcal{V}_{i,j}}\frac{1}{|\mathbf{\nabla}u^{n}|_{\epsilon}}\frac{u^{n+1}-u^{n}}{\tau_{S}} d\mathbf{x}=\int_{\partial \mathcal{V}_{i,j}}g\frac{\mathbf{\nabla}u^{n+1}}{|\mathbf{\nabla}u^{n}|_{\epsilon}}\cdot \mathbf{n}_{i,j}dS,
 \end{equation}
where $\mathbf{n}_{i,j}$ is a unit outward normal vector to the boundary of the pixel $(i,j)$.
In a similar manner as in Equation \ref{average_gradient}, we use the diamond cell approach \cite{mikula2009}. The average of neighboring pixel values in the four corners of the pixel $(i,j)$ are written as

\begin{equation} 
\begin{aligned}
u^{1,1}_{i,j}=\frac{1}{4}(u^{n}_{i,j}+u^{n}_{i,j+1}+u^{n}_{i+1,j}+u^{n}_{i+1,j+1}),\\
u^{1,-1}_{i,j}=\frac{1}{4}(u^{n}_{i,j}+u^{n}_{i+1,j}+u^{n}_{i,j-1}+u^{n}_{i+1,j-1}),\\
u^{-1,-1}_{i,j}=\frac{1}{4}(u^{n}_{i,j}+u^{n}_{i-1,j}+u^{n}_{i,j-1}+u^{n}_{i-1,j-1}), \\
u^{-1,1}_{i,j}=\frac{1}{4}(u^{n}_{i,j}+u^{n}_{i,j+1}+u^{n}_{i-1,j}+u^{n}_{i-1,j+1}). \\
\end{aligned}
 \end{equation}

The gradient of $u^{n}_{i,j}$, in $n^{th}$ step for a pixel $(i,j)$, is approximated by 
\begin{equation} 
\begin{aligned} 
\nabla^{1,0}u^{n}_{i,j}=\frac{1}{h}(u^{n}_{i+1,j}-u^{n}_{i,j},u^{1,1}_{i,j}-u^{1,-1}_{i,j}),\\
\nabla^{0,-1}u^{n}_{i,j}=\frac{1}{h}(u^{1,-1}_{i,j}-u^{-1,-1}_{i,j},u^{n}_{i,j-1}-u^{n}_{i,j}),\\
\nabla^{-1,0}u^{n}_{i,j}=\frac{1}{h}(u^{n}_{i-1,j}-u^{n}_{i,j},u^{-1,1}_{i,j}-u^{-1,-1}_{i,j}),\\
\nabla^{0,1}u^{n}_{i,j}=\frac{1}{h}(u^{1,1}_{i,j}-u^{-1,1}_{i,j},u^{n}_{i,j+1}-u^{n}_{i,j}).\\
\end{aligned}
 \end{equation} 

\noindent
Now we can define
\begin{equation} \label{absolute}
\begin{aligned}
Q_{i,j}^{l,m;n}=\sqrt{\epsilon^2+|\nabla^{l,m}u_{i,j}^{n}|^{2}}\\
\bar{Q}_{i,j}^{l,m;n}=\sqrt{\epsilon^2+\frac{1}{4}\sum_{|l|+|m|=1}|\nabla^{l,m}u_{i,j}^{n}|^{2}},
\end{aligned}
 \end{equation}
\noindent 
where $l,m \in \{-1,0,1\}$, $|l|+|m|=1$, in the set of grid neighbors $N_{i,j}$. 
The final discretized form of Equation \ref{subsurf} is given by \cite{mikula2008}  
 \begin{equation} \label{fin_subsurf}
u^{n+1}_{i,j}-u^{n}_{i,j}=\frac{\tau_{S}}{h^2}\bar{Q}_{i,j}^{l,m;n}\sum_{|l|+|m|=1}g_{i,j}^{l,m,\sigma} \frac{u^{n+1}_{i+l,j+m}-u^{n+1}_{i,j}}{Q_{i,j}^{l,m;n}},
 \end{equation}
where $h^2$ is the pixel area and $g_{i,j}^{l,m,\sigma}=g(|\nabla^{l,m} I^{0}_{i,j;\sigma}|)$.
Equation \ref{fin_subsurf} is solved using the SOR method, and the relaxation factor was set to $1.8$. The calculation was stopped when $\sum_{i=1}^{M}\sum_{j=1}^{N}|u_{i,j}^{n+1}-u_{i,j}^{n}| < 0.01$.

\begin{table}[ht]
    \centering
    \renewcommand{\arraystretch}{1.3}
    \resizebox{\textwidth}{!}{
    \begin{tabular}{ c  l | c  l }
    \toprule
        $t$ & Variable denoting time in PDE & $k$ & Index of the image sequence \\
        $T_{F}$ & Upper limit of the filtering scale & $(i,j)$ & Indices of a pixel position \\
        $\theta$ & Variable denoting  the real time, the sequence of images& $h$ & Pixel size \\
        $\theta_{F}$ & Upper limit of the image sequence & $W_{i,j}$ & A local window centered by a pixel $(i,j)$ \\
        $(x_{1},x_{2})$ & A point in $R^{2}$ & $T_{i,j}$ & A general threshold value in $W_{i,j}$ \\ 
        $u$ & Unknown real function & $T_{i,j}^{*}$ & The optimal threshold in $W_{i,j}$ \\
        $clt(u)$ & Scalar function measuring the temporal coherence of objects & $\sigma^{2}_{B}$ & Variance between classes of the background and objects \\
        $G_{\sigma}$ & Gaussian function with variance $\sigma$  & $I(i,j)$ & Image intensity of the pixel $(i,j)$ \\
        $g$ & Edge detector function &  $\delta$ & Parameter checking if a local window contains objects \\
        $\tau_{F}$ & Step size in space-time filtering & $T_{S}$ & Upper limit of SUBSURF \\
        $n$ & Iteration index in filtering and SUBSURF &  $\tau_{S}$ & Step size in SUBSURF \\ [0.5ex]
    \bottomrule
    \end{tabular}
    }
    \caption{Overview of symbols used in the proposed segmentation.}
    \label{table_segmentation}
\end{table}

\subsection{Extraction of macrophage trajectories}
\subsubsection{Detection of the approximate cell center}
This section describes the time-relaxed eikonal equation employed to find the cell centers using segmentation results. 
As shown in Fig. \ref{fig_seg3}, the segmentation does not always extract the whole shape of some macrophages. Therefore, the connected segmented subregions are used in this method, and we will call the connected segmented subregions segmented regions in short.
We approximate the centers of the segmented region by finding the maxima of the distance function evaluated from the boundary of the cells and solved inside the segmented cells.
The computation of the distance function by solving the time-relaxed eikonal equation is restricted only to the area of segmented regions.
This approach guarantees that the centers obtained by using the distance function are always inside any shapes of segmented regions.
Since the center---the local maxima of the distance function---are not identical to the actual cell centers, they will be called ``approximate cell centers''.
In this section, we describe the solution of the eikonal equation by the Rouy--Tourin scheme. The time-relaxed eikonal equation is written as
\begin{equation}
    \frac{\partial d}{\partial t}+|\nabla d|=1.
    \label{eikonal}
\end{equation}
In every time slice $\theta \in [0, \theta_{F}]$, we solve Equation \ref{eikonal} for the unknown function $d(t, \mathbf{x}, \theta)$ where $(t, \mathbf{x}) \in [0,T_{E}] \times \Omega$.
The equation is discretized by the explicit scheme using the step size $\tau_{D}$, and the Rouy--Tourin scheme is used for space discretization \cite{osher1988, rouy1992, sethian1996}. We solve Equation \ref{eikonal} in every 2D data slice. Let $d_{i,j}^{n}(\theta)$ denote the approximate solution of Equation \ref{eikonal} at the time slice $\theta$ in a pixel $(i,j)$ at a discrete step $t^{n}=n\tau_{D}$. For every $(i,j)$, the index set $N_{i,j}$ consists of all $(l,m)$ such that $l,m \in \{-1,0,1\}$, $|l|+|m|=1$, and then $D^{l,m}_{i,j}(\theta)$ is defined for any $(l,m)$ as

\begin{equation}
    D^{l,m}_{i,j}(\theta)=\biggl(\text{min} \biggl(d^{n}_{i+l,j+m}(\theta)-d^{n}_{i,j}(\theta),0\biggr) \biggr)^{2}.
    \label{dis_D}
\end{equation}
In addition, 
\begin{equation}
    \begin{split}
    M^{1,0}_{i,j}(\theta)=\text{max} \biggl(D^{-1,0}_{i,j}(\theta), D^{1,0}_{i,j}(\theta) \biggr),\\
    M^{0,1}_{i,j}(\theta)=\text{max} \biggl(D^{0,-1}_{i,j}(\theta), D^{0,1}_{i,j}(\theta) \biggr),\\
    \end{split}
     \label{dis_M}
\end{equation}
are defined.
Finally, the discretization of Equation \ref{eikonal} at time slice $\theta$ takes the following form,

\begin{equation}
    d^{n+1}_{i,j}(\theta)=d^{n}_{i,j}(\theta)+\tau_{D}-\frac{\tau_{D}}{h}\sqrt{M^{1,0}_{i,j}(\theta)+M^{0,1}_{i,j}(\theta)},
    \label{dis_eikonal}
\end{equation}
where $\tau_{D}=h/2$ is used for stability reasons.
In this paper, this equation is solved only inside the segmented regions according to the following process.
The first step is to set $d^{0}_{i,j}(\theta)=0$ inside the segmented regions and $d^{0}_{i,j}(\theta)=BIG$ outside the segmented regions; here, the value $BIG$ is much greater than $0$. Next, the numerical scheme in Equation \ref{dis_eikonal} is applied only inside the segmented regions. $d_{i,j}^{n}(\theta)$ is fixed to $0$ at the boundary of segmented regions by considering a pixel $(i,j)$ is at the boundary in case that $d^{n}_{i,j}(\theta) \neq BIG$ and there is at least one neighboring pixel which fulfills $d^{n}_{i+l,j+m}(\theta) = BIG$. The computation is stopped when the inequality  $\sum_{\theta=0}^{\theta_{F}}\sum_{i=1}^{M}\sum_{j=1}^{N}|d_{i,j}^{n+1}(\theta)-d_{i,j}^{n}(\theta)| < 0.001$ is fulfilled , and the values of $d_{i,j}^{n+1}(\theta)$ at the last time step is used as $d_{i,j}(\theta)$ in the next section.

\subsubsection{Extraction of partial trajectories} \label{algorithm_partial}
In this section, we introduce an algorithm to connect the approximate cell centers in the case of overlapping macrophages using the backtracking approach. The algorithm yields the trajectories that connect the cells overlapping in the temporal direction---all these trajectories will be called partial trajectories.

Three sets of values will play a major role in the algorithm: $d_{i,j}(\theta)$, $\mathcal{F}_{i,j}(\theta)$, and $C^{l}(\theta)$, where $(i,j)$ denote a pixel position, $\theta$ denotes a time slice and $l$ denotes the cell center number. 

\begin{figure}[ht]
	\centering
   \includegraphics[scale=0.95]{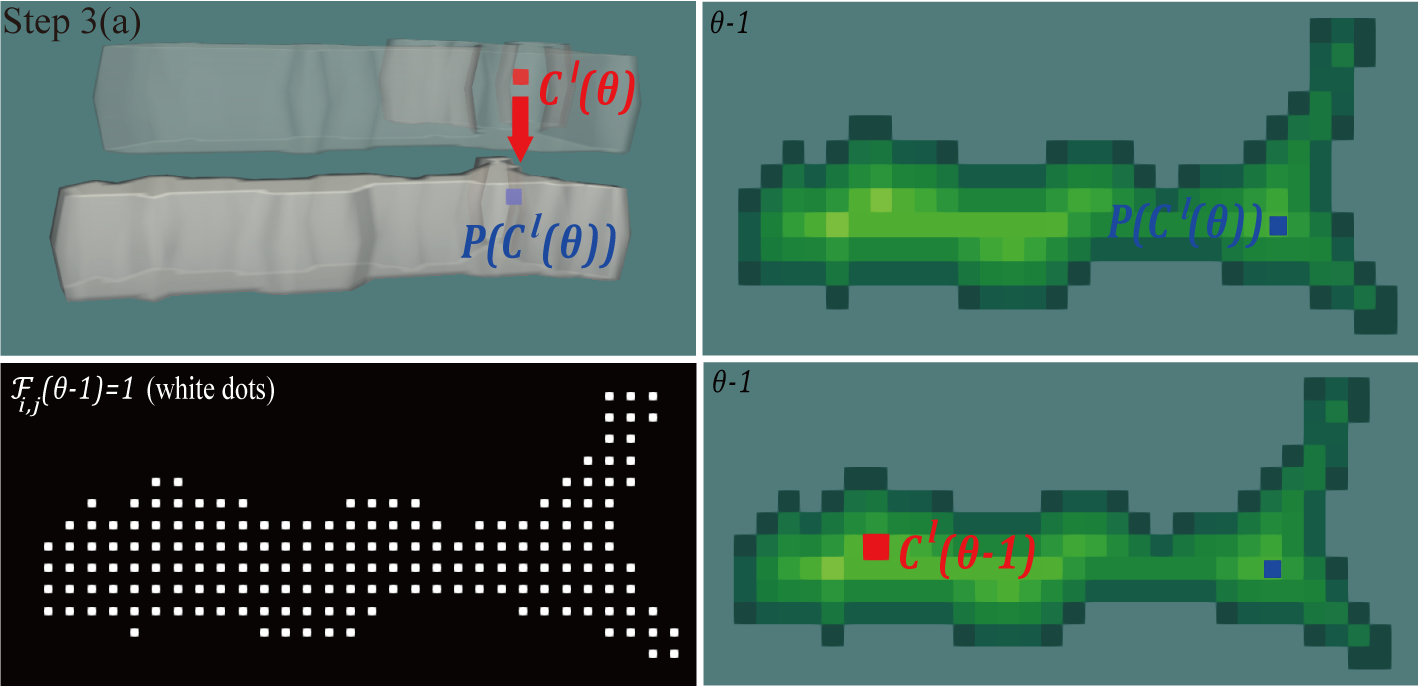}
    \caption{Schematic picture of step 3(a) of the proposed algorithm. The cells in the top-left panel are amplified along the time axis for better visualization. The blue dot denotes the projected coordinate.}
   \label{fig_track_step3a}
\end{figure}

\begin{figure}[ht]
	\centering
   \includegraphics[scale=0.95]{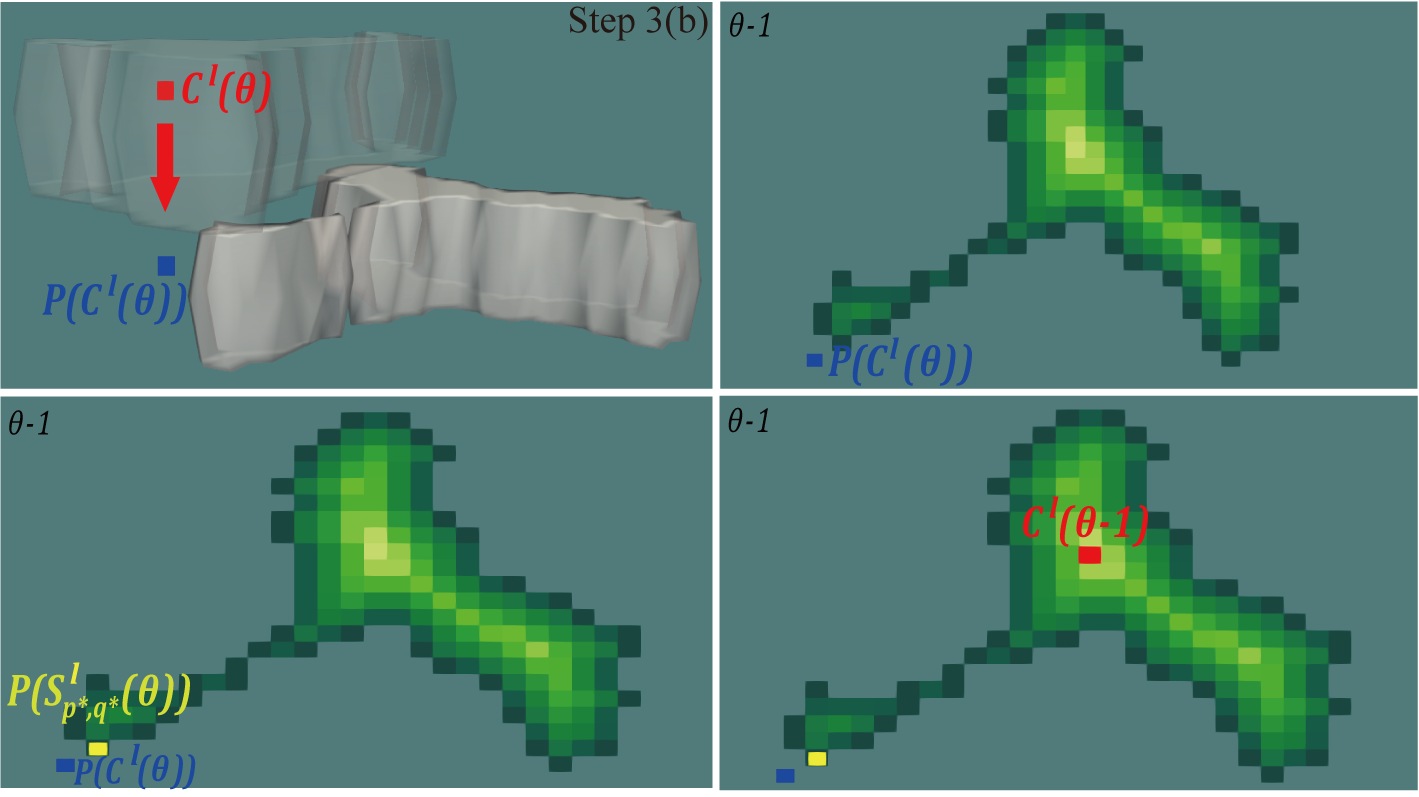}
    \caption{Schematic picture of steps 3(b) of the proposed algorithm. The cells in the top-left panel are amplified along the time axis for better visualization. The yellow dot denotes the inspected coordinate where $d_{ij}(\theta-1) \neq BIG$. Step 3(b)i is shown in the bottom panels.}
   \label{fig_track_step3b}
\end{figure}

\begin{figure}[ht]
	\centering
   \includegraphics[scale=0.55]{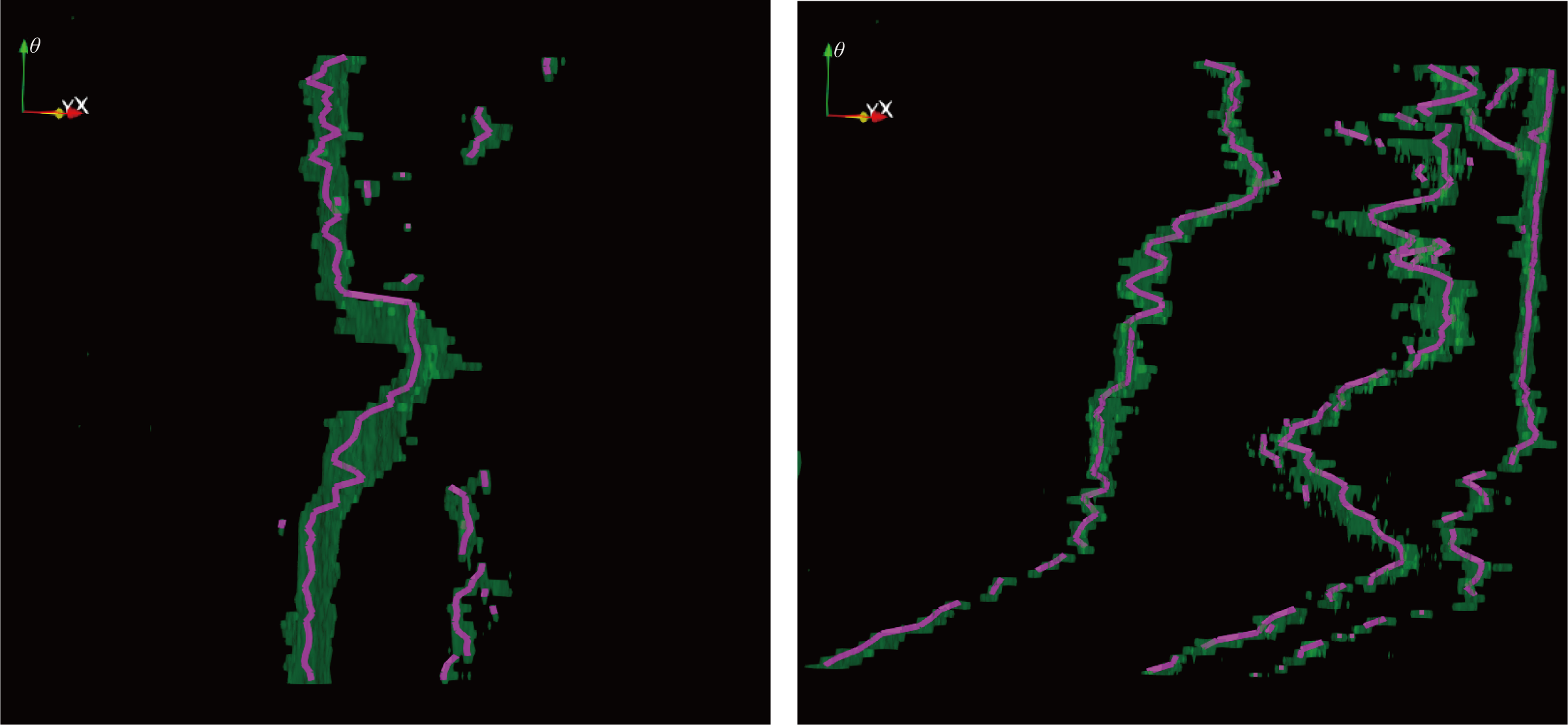}
    \caption{Partial trajectories of several macrophages. The time axis is amplified for better visualization.}
   \label{fig_track_partial}
\end{figure}

First, the distance function value $d_{i,j}(\theta)$ indicates whether a pixel is inside a segmented region or not. The pixel $(i,j)$ at time slice $\theta$ is positioned inside the segmented region if $d_{i,j}(\theta) \neq BIG$. 
Second, $\mathcal{F}_{i,j}(\theta)=1$ indicates the pixel $(i,j)$ belongs to the segmented region which is already connected to another cell by a partial trajectory. 
Lastly, $C^{l}(\theta)$ represents the selected cell center, $l=1,..., N^{\theta}$, where $N^{\theta}$ is the total number of segmented regions at the time step $\theta$. With these definitions, the steps for linking the approximate cell centers are as follows:

\begin{enumerate}
    \item For all pixels $(i,j)$ and all time steps $\theta$,  $\mathcal{F}_{i,j}(\theta)$ is set to $0$ and $d_{i,j}(\theta)$ is computed by the method in Equation \ref{dis_eikonal}. 
    \item Let $\theta_{L}$ be a time slice and let $\theta_{L}=\theta_{F}$ initially. The values of the distance function ($d_{i,j}(\theta_{L}) \neq BIG$) inside every segmented region in time slice $\theta_{L}$  are inspected  and the pixel having the maximal value of distance function inside the segmented region is found and designated as approximate cell center, $C^{l}(\theta_{L})=\bigl(C_{1}^{l}\bigl(\theta_{L}), C_{2}^{l}(\theta_{L})\bigr)$, $l=1,..., N^{\theta}$.
    \item Let $\theta=\theta_{L}$. In a backtracking manner, we look for overlapping segmented regions by performing steps (a)-(b): for $l=1, ..., N^{\theta}$,  
    $C^{l}(\theta)$ is projected onto the spatial plane of the previous time slice $\theta-1$. Let denote the projected point as $P(C^{l}(\theta))$, where $P(C^{l}(\theta))=\bigl(C_{1}^{l}(\theta), C_{2}^{l}(\theta),\theta-1 \bigr)$.
    \begin{enumerate}
    \item \label{ptra1} The case when an approximate cell center is projected inside some segmented region:\mbox{} \\
    If $d_{i,j}(\theta-1) \neq BIG$ for $(i,j)=
    \bigl(C_{1}^{l}(\theta), C_{2}^{l}(\theta)\bigr)$,
    the approximate cell center at $\theta -1$ is found by searching for the maximum value of the distance function inside the segmented region at time $\theta-1$, and the approximate cell center is denoted by $C^{l}(\theta-1)=\bigl(C_{1}^{l}(\theta-1), C_{2}^{l}(\theta-1)\bigr)$. \\
    Also, $\mathcal{F}_{i,j}(\theta-1)$ is changed to $1$ for all pixels $(i,j)$ inside the corresponding segmented region. After finding the approximate cell center $C^{l}(\theta-1)$, it is connected with $C^{l}(\theta)$, forming a section of the partial trajectory (see Fig. \ref{fig_track_step3a}).
    \item \label{checkoverlap}
    The case when the projected cell center is not inside of any segmented region at time $\theta-1$: \mbox{} \\
    If $d_{i,j}(\theta-1) = BIG$ for $(i,j)=
    \bigl(C_{1}^{l}(\theta), C_{2}^{l}(\theta)\bigr)$, let $S^{l}(\theta)$ be a set of all pixels $(i,j)$ belonging to the $l^{th}$ segmented region at time $\theta$. Then $d_{i,j}(\theta-1)$ is inspected for all pixels $(i,j)$ in $S^{l}(\theta)$. The inspection is stopped if $d_{i,j}(\theta-1) \neq BIG$ for some $(i,j)= (p^{*}, q^{*})$ and denoting such point $S_{p^{*}, q^{*}}^{l}(\theta)$, or if all pixels in $S^{l}(\theta)$ are inspected without finding such a point.
        \begin{enumerate}
        \item \label{ptra2} Suppose a point $S_{p^{*}, q^{*}}^{l}(\theta)$ exists. In that case, the approximate cell center $C^{l}(\theta-1)=\bigl(C_{1}^{l}(\theta-1), C_{2}^{l}(\theta-1)\bigr)$ is found like in the step $3(a)$ but starting from $P(S_{p^{*}, q^{*}}^{l}(\theta))$, and $\mathcal{F}_{i,j}(\theta-1)$ is set to $1$ for all pixels inside the segmented region at $\theta-1$ to which $P(S_{p^{*}, q^{*}}^{l}(\theta))$ belongs to. After finding the approximate cell center, $C^{l}(\theta-1)$ is connected with $C^{l}(\theta)$, forming a section of the partial trajectory.
         \item Suppose a point $S_{p^{*}, q^{*}}^{l}(\theta)$ does not exist. In that case, the approximate cell center is not designated because there is no overlap of the segmented region $l$ at $\theta$ with any segmented region at $\theta-1$. 
       \end{enumerate} 
    \end{enumerate}
    \item Step $3$ is repeated by decreasing $\theta$ by one until $\theta =1$.
    \item $\theta_{L}$ is decreased by one and $d_{i,j}(\theta_{L})$ and $\mathcal{F}_{i,j}(\theta_{L})$ are checked for all $(i,j)$.\\
    If there is a pixel that fulfills $d_{i,j}(\theta_{L}) \neq BIG $ and $\mathcal{F}_{i,j}(\theta_{L}) =0$, we consider that the pixel is inside a segmented region at $\theta_{L}$ non-overlapping with segmented regions at $\theta_{L}+1$. The approximate cell center of the region is found like step $2$. Then, the steps $3$ and $4$ are repeated. 
\end{enumerate}
Fig. \ref{fig_track_step3a} depicts the case when the projected cell center is located inside some segmented cell (blue dot), and the approximated cell center at $\theta-1$ has been found as a maximum value of distance function inside the cell at $\theta-1$ (red dot in the bottom-right panel). In the situation when the projected center is outside of any segmented cell, a suitable pixel on the boundary of the cell at $\theta-1$ is found, and step $3(b)$ is performed (Fig. \ref{fig_track_step3b}).
Let us note that in steps 3(a) and 3(b), if there are several maxima of the distance function in the inspected segmented region, then the cell center is chosen as the first one found. In step 3, the trajectories can remain disconnected if there is no overlap of cells and the condition 3(b)ii is fulfilled. In Fig. \ref{fig_track_partial}, such partial trajectories are depicted inside the 3D spatial-temporal structure formed by stacking segmented regions in the temporal direction. This result shows that the algorithm works correctly for the overlapped cells, and the extracted partial trajectories appear as expected.

\newpage
\subsubsection{Connection of partial trajectories} \label{connection}
\begin{figure}[ht]
   \includegraphics[scale=1.0]{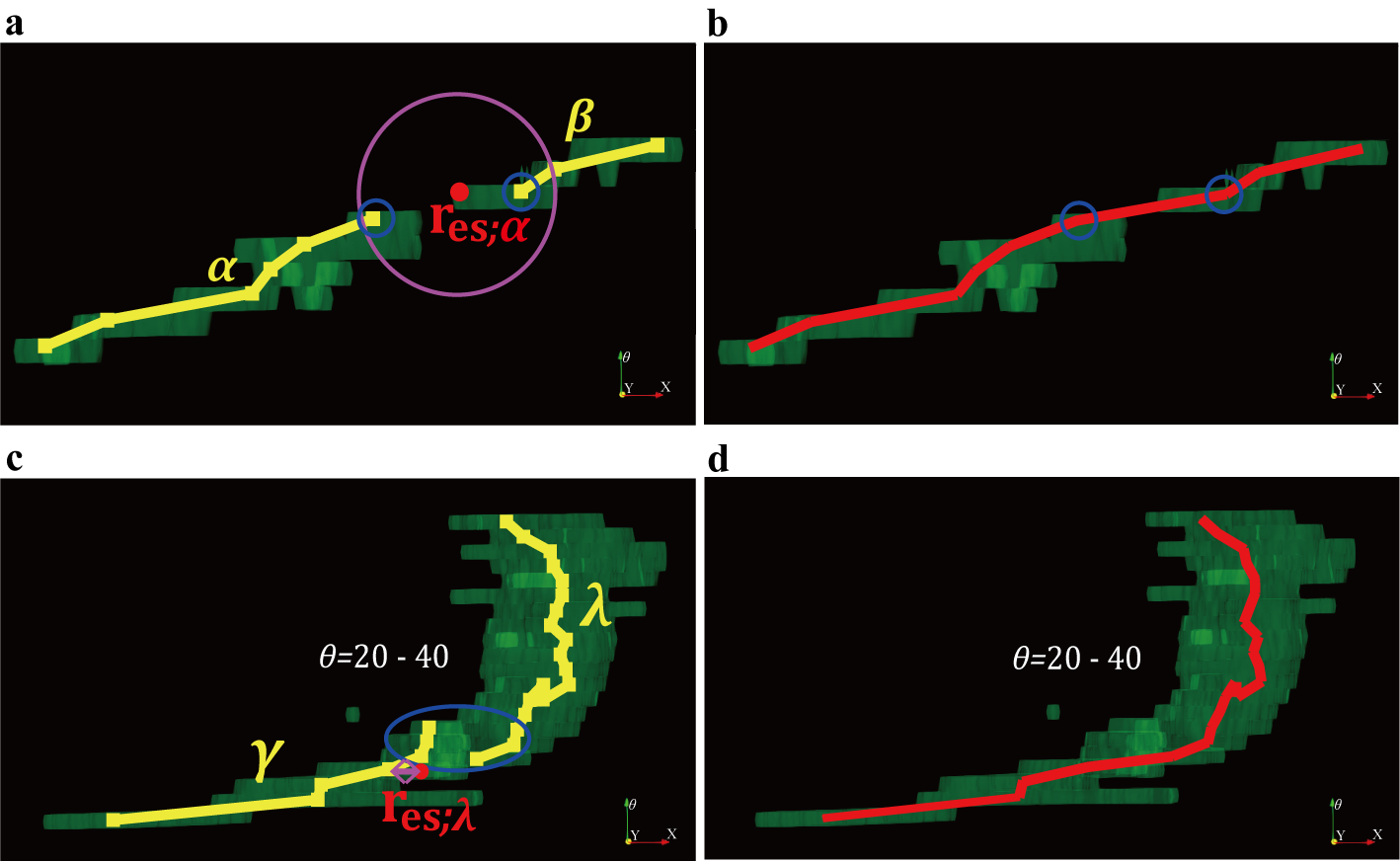}
   \centering
    \caption{\textbf{a}: Two different partial trajectories $\alpha$ and $\beta$. \textbf{b}: the connected trajectory containing $\alpha$ and $\beta$. The blue circles show the last point of the partial trajectory $\alpha$ and the beginning point of $\beta$. \textbf{c}: Two different trajectories $\gamma$ and $\lambda$. \textbf{d}: The connected trajectory containing $\gamma$ and $\lambda$. The blue circle shows the points of the two trajectories in the common time slices.}
   \label{fig_track_connection}
\end{figure}

\begin{figure}[ht]
   \includegraphics[scale=0.6]{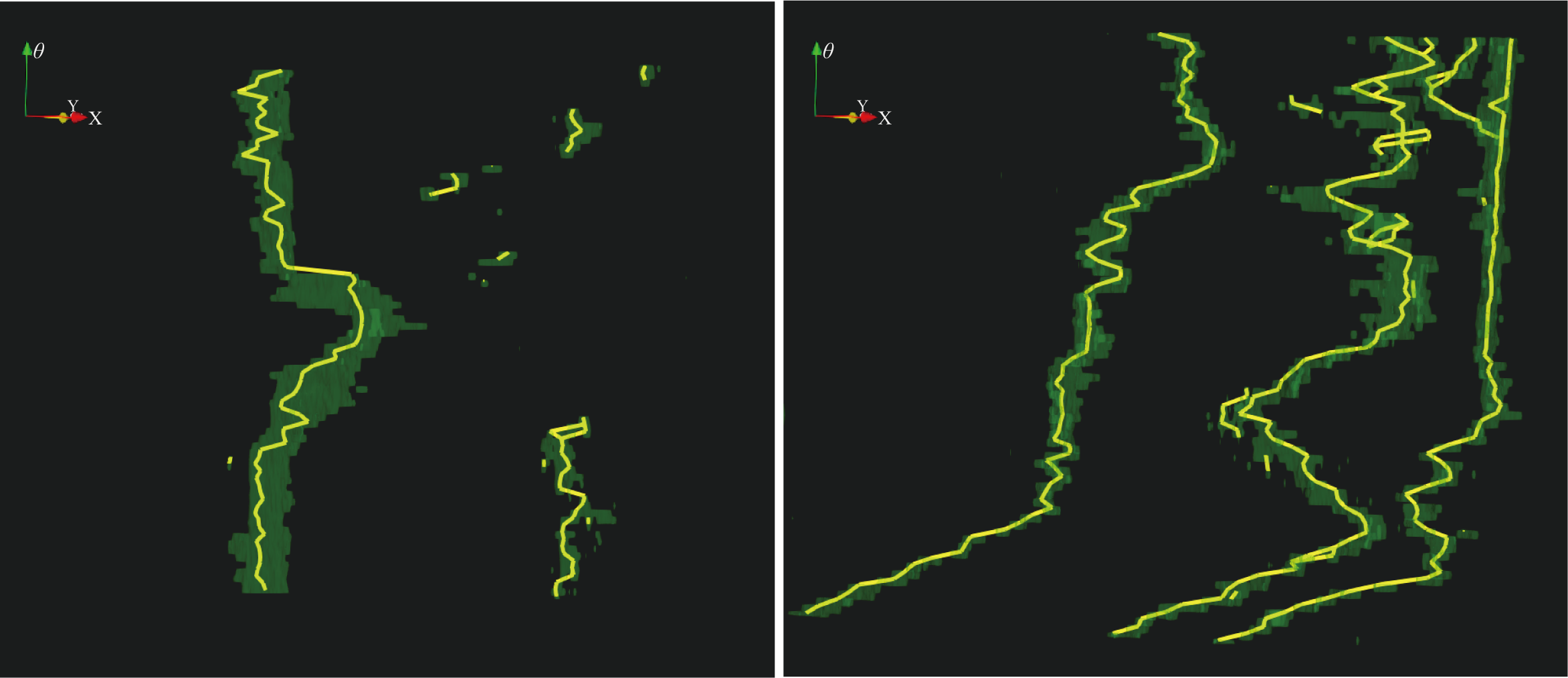}
   \centering
    \caption{Connected trajectories from the partial trajectories in Fig. \ref{fig_track_partial}.}
   \label{fig_track_connected}
\end{figure}

This section describes how the entire trajectories are reconstructed by connecting the partial trajectories. We assume that the reason for macrophages' non-overlap in time is that their movement is relatively fast. It is highly probable that fast-moving macrophages are not imaged continuously in the time since the time step for imaging is fixed. Therefore, we approximate the non-overlapping macrophages in time by keeping their direction of movement. In other words, we expect that the tangents at the endpoints of the partial trajectory are similar to those of corresponding macrophages at the next/previous time slice.
\newline
\indent The tangent approximation by the backward finite difference is used to estimate the position of a point in the next time step of a partial trajectory. Similarly, the forward difference for the tangent calculation is used to estimate the point at the previous time step of the partial trajectory. In the tangent calculation, third-order accuracy is maximally considered, and thus there are three forms of tangent approximation depending on the number of points in the partial trajectory.
The tangents computed with the third order accuracy using the backward and forward finite difference approximations are given by \cite{fornberg1988}

\begin{equation}
    \begin{split}
    V^{b}(\textbf{r}_{\theta})=\frac{1}{\Delta \theta}\left(\frac{11}{6}\textbf{r}_{\theta}-3\textbf{r}_{\theta-1}+\frac{3}{2}\textbf{r}_{\theta-2}-\frac{1}{3}\textbf{r}_{\theta-3}\right),\\\\
    V^{f}(\textbf{r}_{\theta})=\frac{1}{\Delta \theta}\left(-\frac{11}{6}\textbf{r}_{\theta}+3\textbf{r}_{\theta+1}-\frac{3}{2}\textbf{r}_{\theta+2}+\frac{1}{3}\textbf{r}_{\theta+3}\right), \\\\
    \end{split}
     \label{tangent_3rd}
\end{equation}
where $\Delta \theta$ means the size of the time slice difference, and $\textbf{r}_{\theta}=(x_{\theta},y_{\theta})$ is the point of the partial trajectory in the time slice $\theta$. In a similar way, second and first order accuracy approximations for backward and forward finite differences are given by 
\begin{equation}
    \begin{split}
   V^{b}(\textbf{r}_{\theta})=\frac{1}{\Delta \theta}\left(\frac{3}{2}\textbf{r}_{\theta}-2\textbf{r}_{\theta-1}+\frac{1}{2}\textbf{r}_{\theta-2}\right),\\\\
    V^{f}(\textbf{r}_{\theta})=\frac{1}{\Delta \theta}\left(-\frac{3}{2}\textbf{r}_{\theta}+2\textbf{r}_{\theta+1}-\frac{1}{2}\textbf{r}_{\theta+2}\right),\\\\
    \end{split}
     \label{tangent_2nd}
\end{equation}

and
\begin{equation}
    \begin{split}
        V^{b}(\textbf{r}_{\theta})=\frac{1}{\Delta \theta}\left(\textbf{r}_{\theta}-\textbf{r}_{\theta-1}\right),\\\\
        V^{f}(\textbf{r}_{\theta})=\frac{1}{\Delta \theta}\left(-\textbf{r}_{\theta}+\textbf{r}_{\theta+1}\right). \\\\
    \end{split}
     \label{tangent_1st}
\end{equation}
Let us consider a partial trajectory with a time step range $[a\Delta\theta, b\Delta\theta]$, $a,b$ integers, and denote the positions of the cell center at $a\Delta\theta$ and $b\Delta\theta$ by $\textbf{r}_{a}$ and  $\textbf{r}_{b}$, respectively. Then, the position of the cell center at $(a-1)\Delta\theta$ can be estimated from the tangent at time step $a\Delta\theta$. For example, if the partial trajectory contains more than three points, the tangent obtained by using the forward difference at $a\Delta\theta$ is given by
\begin{equation}
V^{f}(\textbf{r}_{a})=\frac{1}{\Delta \theta}\left(-\frac{11}{6}\textbf{r}_{a}+3\textbf{r}_{a+1}-\frac{3}{2}\textbf{r}_{a+2}+\frac{1}{3}\textbf{r}_{a+3}\right)
\label{tangent_a}
\end{equation}
and the tangent at $(a-1)\Delta\theta$ would be
\begin{equation} 
V^{f}(\textbf{r}_{a-1})=\frac{1}{\Delta \theta}\left(-\frac{11}{6}\textbf{r}_{a-1}+3\textbf{r}_{a}-\frac{3}{2}\textbf{r}_{a+1}+\frac{1}{3}\textbf{r}_{a+2}\right).
\label{tangent_a1}
\end{equation} 
Assuming $V^{f}(\textbf{r}_{a-1}) = V^{f}(\textbf{r}_{a})$, i.e., the uniform directional motion of non-overlapping macrophages, we see that $\textbf{r}_{a-1}$ is the only unknown in the equation and can be easily obtained. Similarly, the tangent at time step $b\Delta\theta$ yields the estimated cell center at time step $(b+1)\Delta\theta$ using the backward finite difference. Therefore, the estimated points on trajectories in the time slice without overlap of cells are given by
\begin{equation}
\begin{aligned}
 \textbf{r}_{a-1} &= -\frac{6}{11}V^{f}(\textbf{r}_{a})\cdot \Delta \theta+\frac{18}{11}\textbf{r}_{a}-\frac{9}{11}\textbf{r}_{a+1}+\frac{2}{11}\textbf{r}_{a+2} , \quad b-a > 2,\\\\
    \textbf{r}_{a-1} &= -\frac{2}{3}V^{f}(\textbf{r}_{a})\cdot \Delta \theta+\frac{4}{3}\textbf{r}_{a}-\frac{1}{3}\textbf{r}_{a+1}, \quad  b-a =2, \\\\
    \textbf{r}_{a-1} &= -V^{f}(\textbf{r}_{a})\cdot \Delta \theta+\textbf{r}_{a}, \quad  b-a =1, \\\\
\end{aligned}
     \label{estPoint_forw}
\end{equation}
and
\begin{equation}
\begin{aligned}
 \textbf{r}_{b+1}&=\frac{6}{11}V^{b}(\textbf{r}_{b})\cdot \Delta \theta+\frac{18}{11}\textbf{r}_{b}-\frac{9}{11}\textbf{r}_{b-1}+\frac{2}{11}\textbf{r}_{b-2} \quad , b-a > 2,\\\\
    \textbf{r}_{b+1}&=\frac{2}{3}V^{b}(\textbf{r}_{b})\cdot \Delta \theta+\frac{4}{3}\textbf{r}_{b}-\frac{1}{3}\textbf{r}_{b-1} \quad , b-a =2, \\\\
   \textbf{r}_{b+1}&=V^{b}(\textbf{r}_{b})\cdot \Delta \theta+\textbf{r}_{b} \quad , b-a =1. \\\\
\end{aligned}
     \label{estPoint_bakw}
\end{equation}
The connection of partial trajectories is carried out when the estimated cell center $\textbf{r}_{\text{es}}$ given by $\textbf{r}_{a-1}$ or $\textbf{r}_{b+1}$ in Equations \ref{estPoint_forw} or \ref{estPoint_bakw}, is positioned near the endpoint $\textbf{r}_{\text{e}}$ of some existing partial trajectory ending at time slice $(a-1)\Delta\theta$ or starting at $(b+1)\Delta\theta$. It means we check the condition
\begin{equation}
\begin{aligned}
|\textbf{r}_{\text{es}}-\textbf{r}_{\text{e}}| \leq \Delta r
\end{aligned}
     \label{cond_connect}
\end{equation}
where $\Delta r$ is a parameter, and if it is fulfilled, then the partial trajectories are connected. Fig. \ref{fig_track_connection}\textbf{a} shows two partial trajectories denoted by $\alpha$ and $\beta$. The red dot in the figure represents the estimated cell center $\textbf{r}_{\text{es}};\alpha$ computed from the $\alpha$ trajectory with the backward finite difference approximation. The $\alpha$ and $\beta$ trajectories are connected if the beginning point of the $\beta$ trajectory and the estimated cell center from the $\alpha$ trajectory are located within the neighborhood $\Delta r$; Fig. \ref{fig_track_connection}\textbf{b} shows the connected trajectory in such case. The condition (\ref{cond_connect}) is written for the case 
when the difference of time slices between endpoints of partial trajectories equals $1$. However, the partial trajectories are connected similarly when the difference of time slices equals $2$ if two estimated points obtained from two partial trajectories (one in a forward manner and one in a backward way) are located in the same time slice and within the $\Delta r$ neighborhood. 
The connection of partial trajectories using the above approach is shown in Fig. \ref{fig_track_connected}.

Furthermore, the tangent calculation is also used to connect the partial trajectories if the points at the beginning or ending parts of trajectories are located close to each other in several time slices. It can happen if the segmentation of a single macrophage contains several fractions in a few time slices. Fig. \ref{fig_track_connection}\textbf{c} shows two such partial trajectories $\gamma$ and $\lambda$. As shown in the blue circle, there are three common time slices where $\gamma$ and $\lambda$ have trajectory points close to each other. To connect those kinds of partial trajectories, we again calculate the estimated point of the partial trajectory using the tangent approximation and check if there is a point $\textbf{r}_{\text{j}}$ of another trajectory in a close neighborhood of the estimated point.
If yes, then also a difference between the time slice of $\textbf{r}_{\text{j}}$ and the time slice of the endpoint $\textbf{r}_{\text{e}}$ of its trajectory is checked. In other words, we check the number of common time slices $\Theta_{c}$ where two close trajectories appear simultaneously. For instance, in the case of Fig. \ref{fig_track_connection}\textbf{c}, $\Theta_{c}=3$.
Finally, two trajectories are connected if the following two conditions are fulfilled: $|\textbf{r}_{\text{es}}-\textbf{r}_{\text{j}}| \leq \Delta r_{2}$ and $\Theta_{c} \leq \Delta r_{\theta}$. The line in Fig. \ref{fig_track_connection}\textbf{d} shows the connected trajectory.

For the choice of parameters $\Delta r$ and $\Delta r_{2}$, we suggest considering the approximate size of macrophages.
The estimated center of the cell $\textbf{r}_{\text{es}}$ locates near $\textbf{r}_{\text{e}}$ if non-overlapping macrophages keep the direction of movement. Then, the distance between $\textbf{r}_{\text{e}}$ and $\textbf{r}_{\text{es}}$ is less than or equal to the radius of macrophages since $\textbf{r}_{\text{e}}$ stands for the approximate center of the macrophage.
Thus, $\Delta r$ can be chosen proportionally to the radius of macrophages.
On the other hand, $\Delta r_{2}$ can be chosen proportionally to the diameter of macrophages. In this step, we mainly link disconnected trajectories caused by fractions in segmentation. The two centers of the segmentation fractions can be located end to end in the same macrophage.
The choice of $\Delta r_{\theta}$ depends on how many times fractions appear in consecutive time slices. In most situations, fractions caused by the segmentation show one big part and some small parts. The trajectories belonging to the small parts may disconnect soon because they are hard to overlap with macrophages in the previous time slice. Therefore, $\Delta r_{\theta}$ is not necessarily to be large, for instance, the value of $\Delta r_{\theta}$ from 5 to 8 can mostly cover the situation in Fig. \ref{fig_track_connection}\textbf{c}.

\begin{table}[t]
    \centering
    \renewcommand{\arraystretch}{1.3}
    \resizebox{\textwidth}{!}{
    \begin{tabular}{ c l }
    \toprule
    $T_{E}$ & Upper limit of scale in Eikonal equation \\
    $d_{i,j}(\theta)$ & Value of the distance function in a pixel $(i,j)$  at $\theta^{th}$ time slice\\   
    $\mathcal{F}_{i,j}(\theta)$ & Indicator if a segmented region belonging to a pixel $(i,j)$ already formed a partial trajectory  \\   
    $C^{l}(\theta)$ & The approximate center of the $l^{th}$ segmented region at $\theta^{th}$ time slice \\   
    $V^{b}$ & Tangent of a partial trajectory computed by backward finite difference approximation  \\ 
    $V^{f}$ & Tangent of a partial trajectory computed by forward finite difference approximation \\ 
    $\textbf{\textrm{r}}_{\textrm{es}}$ & Estimated cell center \\   
    $\textbf{\textrm{r}}_{\textrm{e}}$ & Endpoint of another partial trajectory \\ 
    $ \Delta r$ & Parameter checking closeness with other partial trajectories\\  
    $\textbf{\textrm{r}}_{\textrm{j}}$ & Endpoint of another partial trajectory obtained after connections from Equation \ref{cond_connect}\\  
    $ \Delta r_{2}$ & Parameter checking closeness with other partial trajectories obtained after connections from Equation \ref{cond_connect}\\  
    $\Theta_{c}$ & The number of common time slices of two close trajectories \\ 
    $\Delta r_{\theta}$ & Parameter checking if two close trajectories correspond to the same macrophage \\ [0.5ex]
    \bottomrule
    \end{tabular}
    }
    \caption{Overview of symbols used in the proposed macrophage tracking.}
    \label{table_tracking}
\end{table}
\begin{center}

\end{center}

\newpage
\section{Results}
\subsection{Visual and quantitative assessment of segmentation}
\begin{figure}[ht]
	\centering
   \includegraphics[scale=0.8]{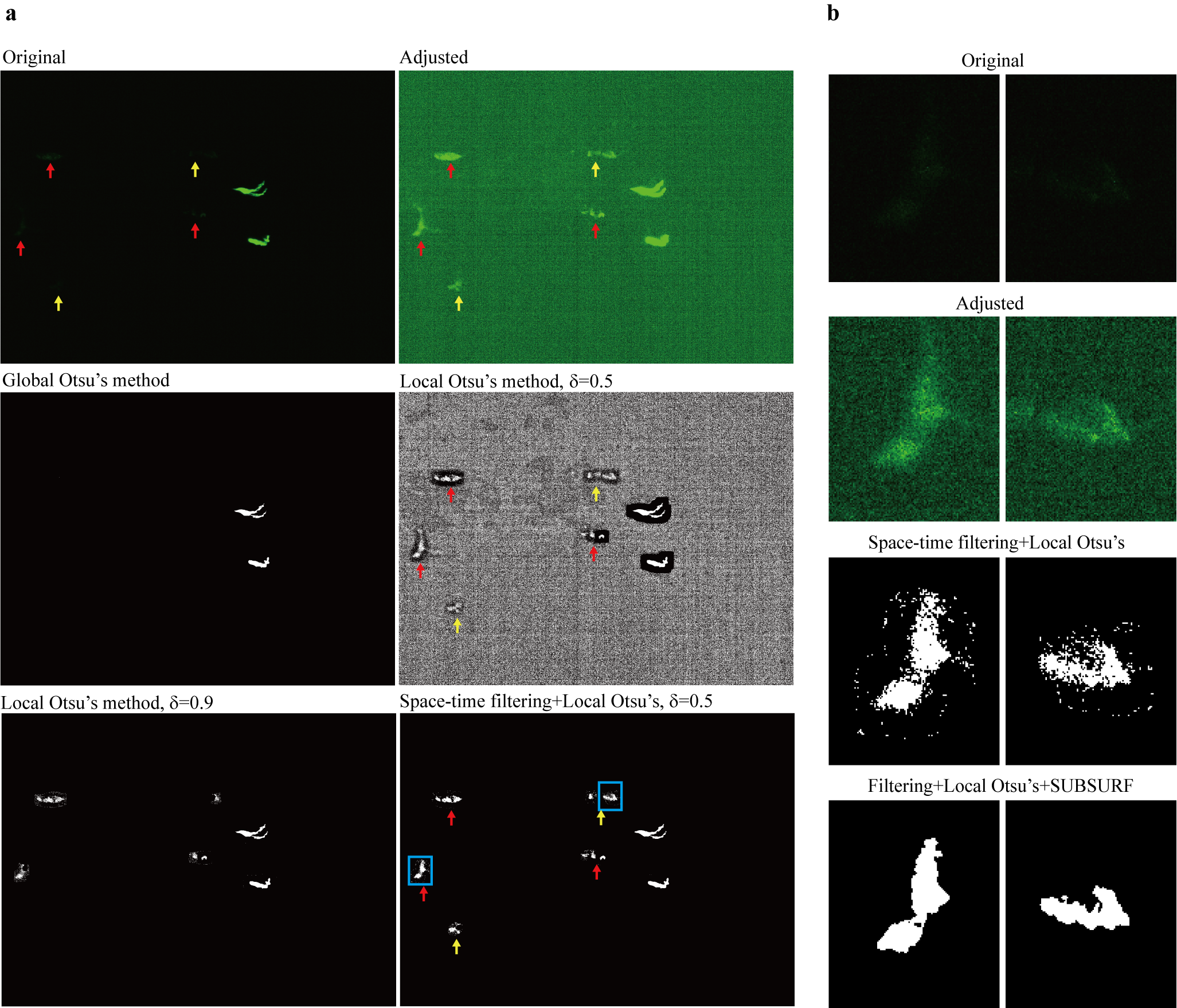}
    \caption{Original and segmented images at $\theta =0$. \textbf{a}: The top-left panel shows the original, and the top-right panel shows after the brightness and contrast of the original images are automatically adjusted. The rest of the panels show the results of applying the global Otsu's method, the local Otsu's method with two different $\delta$, and the local Otsu's method preceded by the space-time filtering. \textbf{b}: Two different macrophages and their processing by the proposed segmentation method. The first row shows the original images where the macrophages are hardly recognizable.
    In the second row, the brightness and contrast of the original images are automatically adjusted. The third row shows already recognizable macrophages in the images obtained by the local Otsu's method with $\delta=0.5$ preceded by the space-time filtering. The last row gives the results after the last segmentation step, the application of the SUBSURF method.}
   \label{fig_proposed_segs}
\end{figure}

We applied the described segmentation method to the second dataset, where macrophages have huge variability of the image intensity.
For the first dataset, we applied the method from \cite{park2020}, using the combination of global thresholding and the SUBSURF method since the macrophages are easily distinguishable from the background due to the relatively weak background noise.
\newline
\indent
The parameters $\tau_{F}$, $K$, $\sigma$ in space-time filtering, $s$, $\delta$ in the local Otsu's method, and $\tau_{S}$, $K$, $\sigma$ in the SUBSURF method are chosen by the parameter optimization (Appendix A).
In Fig. \ref{fig_proposed_segs}\textbf{a}, the images at the time moment $\theta=0$ are shown. The top-left panel shows the original images, and their brightness and contrast are automatically adjusted by using Fiji \cite{schindelin2012fiji} as shown in the top-right panel. 
In the second to the third row, the global and the local Otsu's methods are applied to the original image with different values of parameter $\delta$.
The right panel in the third row shows the result of the local Otsu's method with the filtered images obtained from space-time filtering.
The red and yellow arrows in Fig. \ref{fig_proposed_segs}\textbf{a} show that the global Otsu's method cannot extract those macrophages, but the local Otsu's method can detect and segment them. The ones denoted by the red arrows can be recognizable using both values of $\delta$.
The local Otsu's method without space-time filtering detects the background noise in case of smaller $\delta$ because it captures local information more sensitively.
To avoid the noise from being detected, it can be an option to increase $\delta$; however, macrophages with feeble image intensity cannot be detected. The macrophages denoted by the yellow arrows in Fig. \ref{fig_proposed_segs}\textbf{a} show they are not recognizable when $\delta = 0.9$.
Therefore, filtering is needed before applying the local Otsu's method when images are noisy, and every macrophage with a high variability of image intensity should be detected.
Finally, the background noise disappears when the local Otsu's method with $\delta=0.5$ is applied to the filtered images obtained by space-time filtering. It indicates that space-time filtering makes macrophages distinguishable from the background.
Here, the size of the local window is $50*50$ and the parameters in Equation \ref{final_clt} for these computations were chosen as $\tau_{F}=0.25$, $K=100$, $\sigma=0.1$, and $h=0.1$.
\newline
\indent
To see more details, Fig. \ref{fig_proposed_segs}\textbf{b} shows two macrophages from the ones indicated by red and yellow arrows in Fig. \ref{fig_proposed_segs}\textbf{a}. 
In the second row, the adjusted images are shown to see the shape of the macrophages.
The third row of Fig. \ref{fig_proposed_segs}\textbf{b} shows that the local Otsu's method preceded by the space-time filtering allows the detection of approximate macrophage shapes also in these cases.
However, some black pixels are apparent inside the macrophage shapes since the image intensity of the macrophages is not uniform. In addition, the local window causes an artifact in the form of white pixels around the macrophage. This happens when the local window contains a small part of the macrophage, so $\delta$ has a rather high value. To account for these issues, we use the SUBSURF method, which eliminates the artifacts and smoothes the boundary and interior of the macrophage shapes. The initial level-set function of the SUBSURF method is set to the binary images after the step of local Otsu thresholding. Although the SUBSURF cannot connect the partial fragments of all macrophages, the problems described above are sufficiently solved, see the last row of Fig. \ref{fig_proposed_segs}\textbf{b}. The parameters for the SUBSURF method in Equation \ref{fin_subsurf} for these computations were chosen as $\tau_{S}=0.25$, $\epsilon^{2}=10^{-8}$, $K=10$, $\sigma=1$, and $h=1$.
The suggested steps for macrophage segmentation work reliably for differently shaped macrophages, no matter how complicated their boundaries are.
However, parts of weak image intensity inside macrophages are observed, especially when macrophages stretch their bodies. It yields fractions of segmented regions for a macrophage since the local Otsu's method works locally, and SUBSURF fails to connect the fractions often.

\begin{figure}[ht]
	\centering
   \includegraphics[scale=1.0]{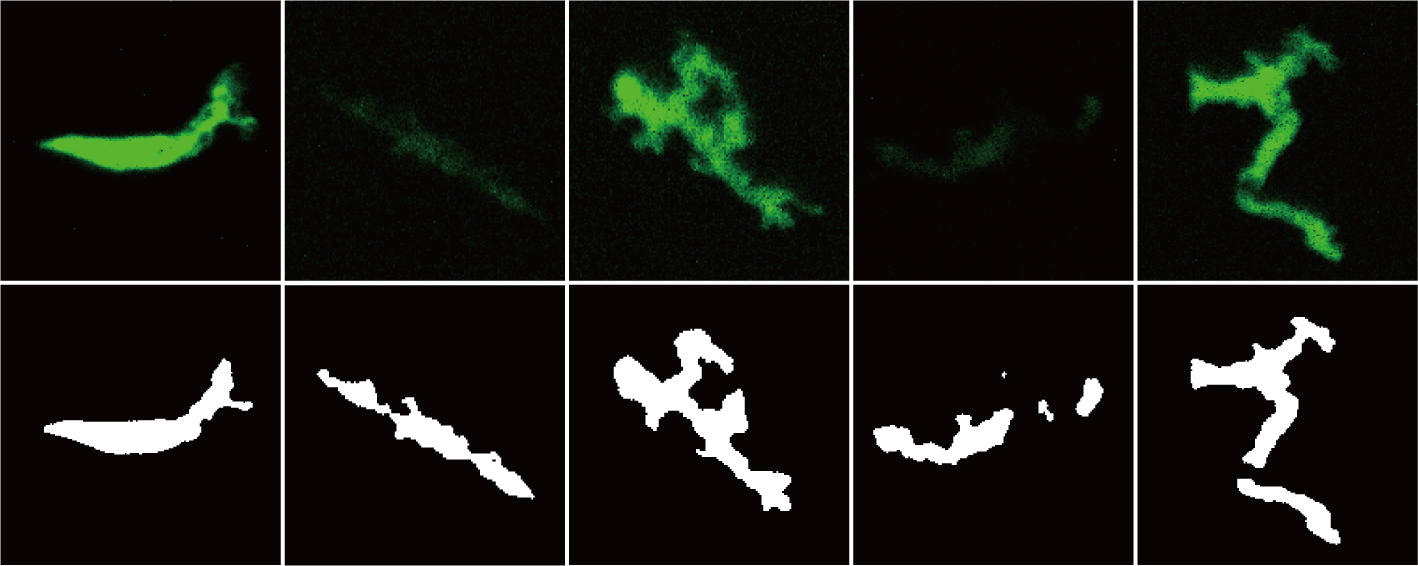}
    \caption{Five different macrophages from the original (top) and the segmented images (bottom). The fourth and fifth columns show the segmentation yields a few segmented regions for a single macrophage.}
   \label{fig_seg3}
\end{figure}

The performance of the presented segmentation method is evaluated quantitatively by using the mean Hausdorff distance of automatic and semi-automatic segmentation results. The mean Hausdorff distance is used to measure how two curves match each other. For two curves given by sets of points, $A=\{a_{1}, ..., a_{N}\}$ and $B=\{b_{1}, ..., b_{M}\}$, the mean Hausdorff distance $d_{\text{H}}$ is defined \cite{mikula2021automated} as $d_{\text{H}}=(\bar{\delta}_{H}(A,B)+\bar{\delta}_{H}(B,A))/2$, where $\bar{\delta}_{H}(A,B)$ is defined as $\bar{\delta}_{H}(A,B)=\frac{1}{M}\sum_{i=1}^{M}\min_{a_{j} \in A}(d_{e}(a_{j},b_{i}))$ . Here, $d_{e}(a_{j},b_{i})$ is the Euclidean distance between $a_{j}$ and $b_{i}$. 
The boundaries of the segmented regions from the automatic and semi-automatic methods are extracted for this. 
The semi-automatic segmentation method, based on the Lagrangian approach \cite{mikula2018}, is done by an expert to create the ``gold standard'' for comparison, see also \cite{park2020}. For the quantitative comparison, we choose two macrophages (the first and fifth macrophages in Fig. \ref{fig_seg3}). They move and change their shapes, covering high variability of segmented shapes over the number of time slices in 120 and 107, respectively. 
The perimeter, area, and circularity $(4\pi*\text{area/perimeter}^{2})$ are calculated for both automatic and semi-automatic segmentation, as shown in Fig. \ref{fig_seg_quantitative1}\textbf{d}--\textbf{f} and Fig. \ref{fig_seg_quantitative2}\textbf{d}--\textbf{f}.
In Fig. \ref{fig_seg_quantitative1}\textbf{a}, the mean Hausdorff distance $d_{\text{H}}$ (measured in pixel units) of the curves representing results of automatic and semi-automatic segmentations is presented over time. 
Also, the quality of the segmentation is measured by using the IoU (Jaccard) index \cite{jaccard1912distribution} and the Sørensen--Dice coefficient \cite{dice1945measures,sorensen1948method}.
For each of the two macrophages, the mean Hausdorff distance, IoU index, and Sørensen--Dice coefficient are averaged over time as shown in Table \ref{table_segment_Hd}.
We see that the average of the mean Hausdorff distances for two macrophages is small compared to the size of macrophages. The IoU index and Sørensen--Dice coefficient obtained from the proposed method indicate the results show reasonable performance.
From Fig. \ref{fig_seg_quantitative1} and Fig. \ref{fig_seg_quantitative2}, we also see that the area of segmented macrophages obtained from the automatic segmentation is slightly smaller than the area obtained by the semi-automatic segmentation. The reason is that the automatic segmentation method does not always detect all parts of the macrophage (fourth and fifth column in Fig. \ref{fig_seg3}) or may give its a more narrow shape.
On the other hand, the presented segmentation method is able to detect every macrophage, although sometimes only partially, which is beneficial for accurate tracking of all macrophages in time-lapse data.

\begin{figure}[ht]
	\centering
   \includegraphics[scale=0.85]{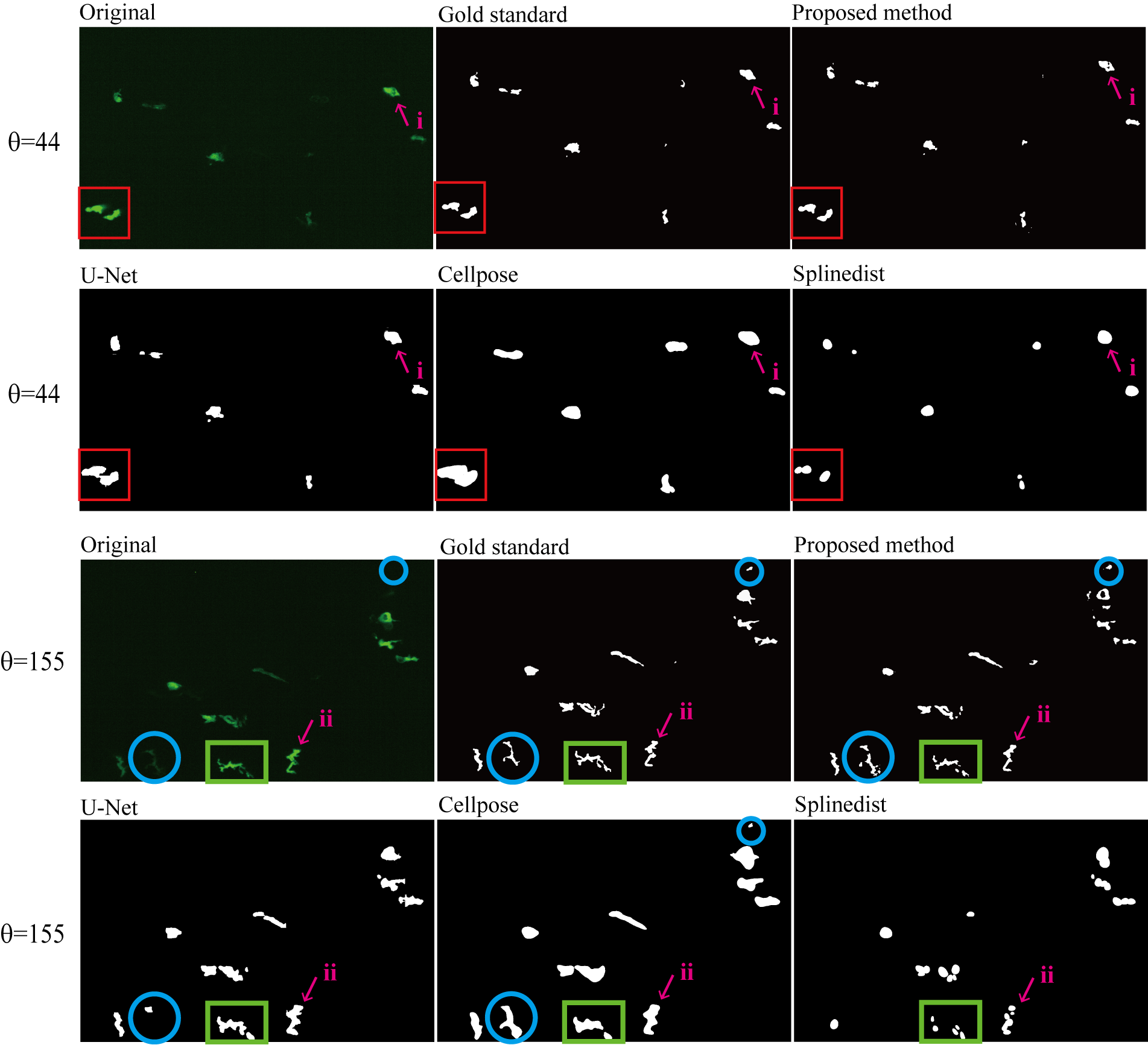}
    \caption{The original, the gold standard images, and segmented images in two different time moments, $\theta=44$ (first--second row) and $\theta=155$ (third--fourth row). The segmentation method used is indicated in the top-left corners of each panel.}
   \label{fig_seg_MLcompare}
\end{figure}

\begin{figure}[ht]
	\centering
   \includegraphics[scale=0.42]{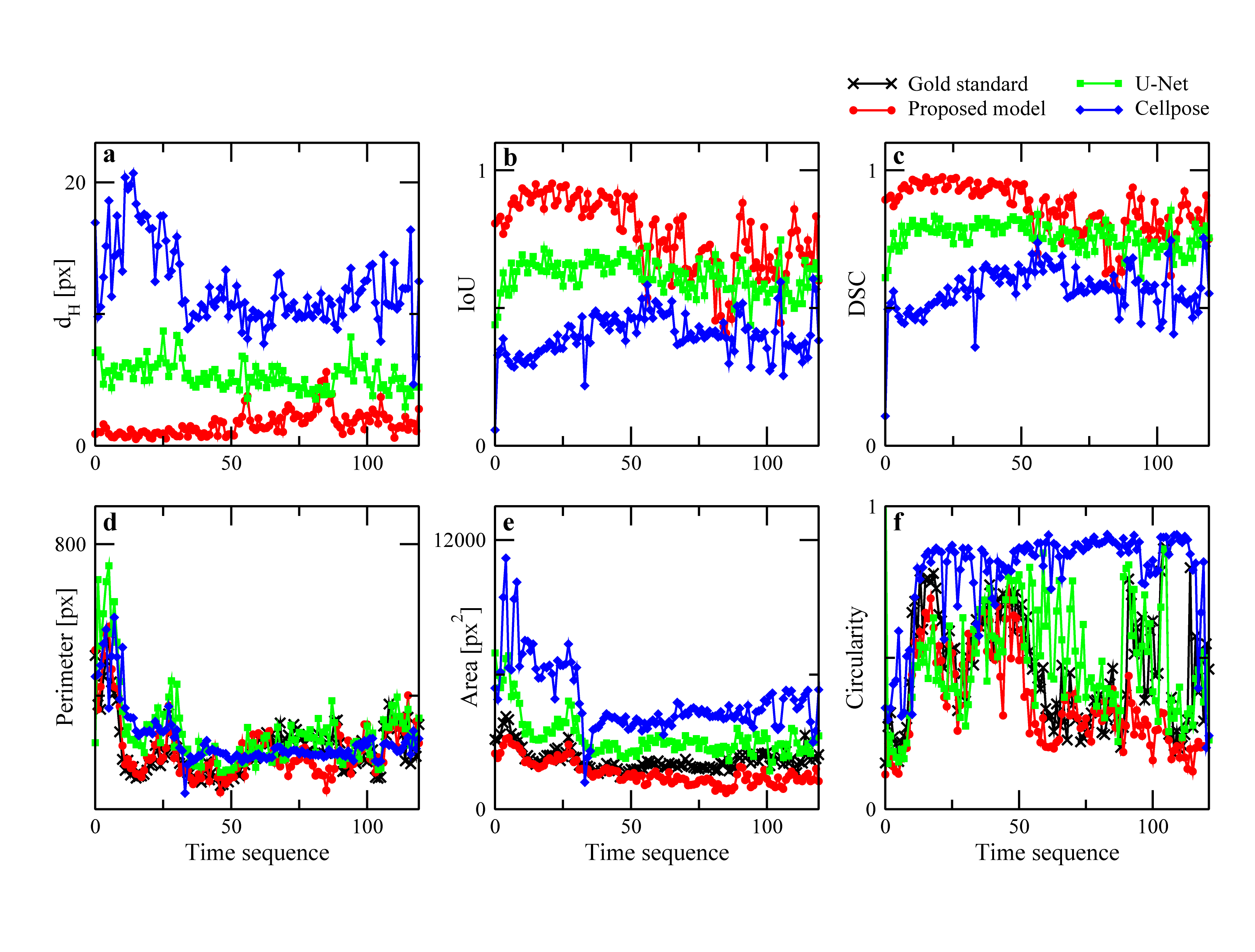}
    \caption{The quantitative comparison for the macrophage denoted by ``i'' in Fig. \ref{fig_seg_MLcompare}. \textbf{a}: The mean Hausdorff distance from the gold standard, \textbf{b}: IoU (Jaccard) index, \textbf{c}: Sørensen--Dice coefficient, \textbf{d}: Perimeters of segmentations, \textbf{e}: Areas of segmentations, and their \textbf{f}: circularities. Time sequence in the horizontal axis indicates the order of time frames, and its interval is 4 minutes.}
   \label{fig_seg_quantitative1}
\end{figure}

\begin{figure}[ht]
	\centering
    \includegraphics[scale=0.42]{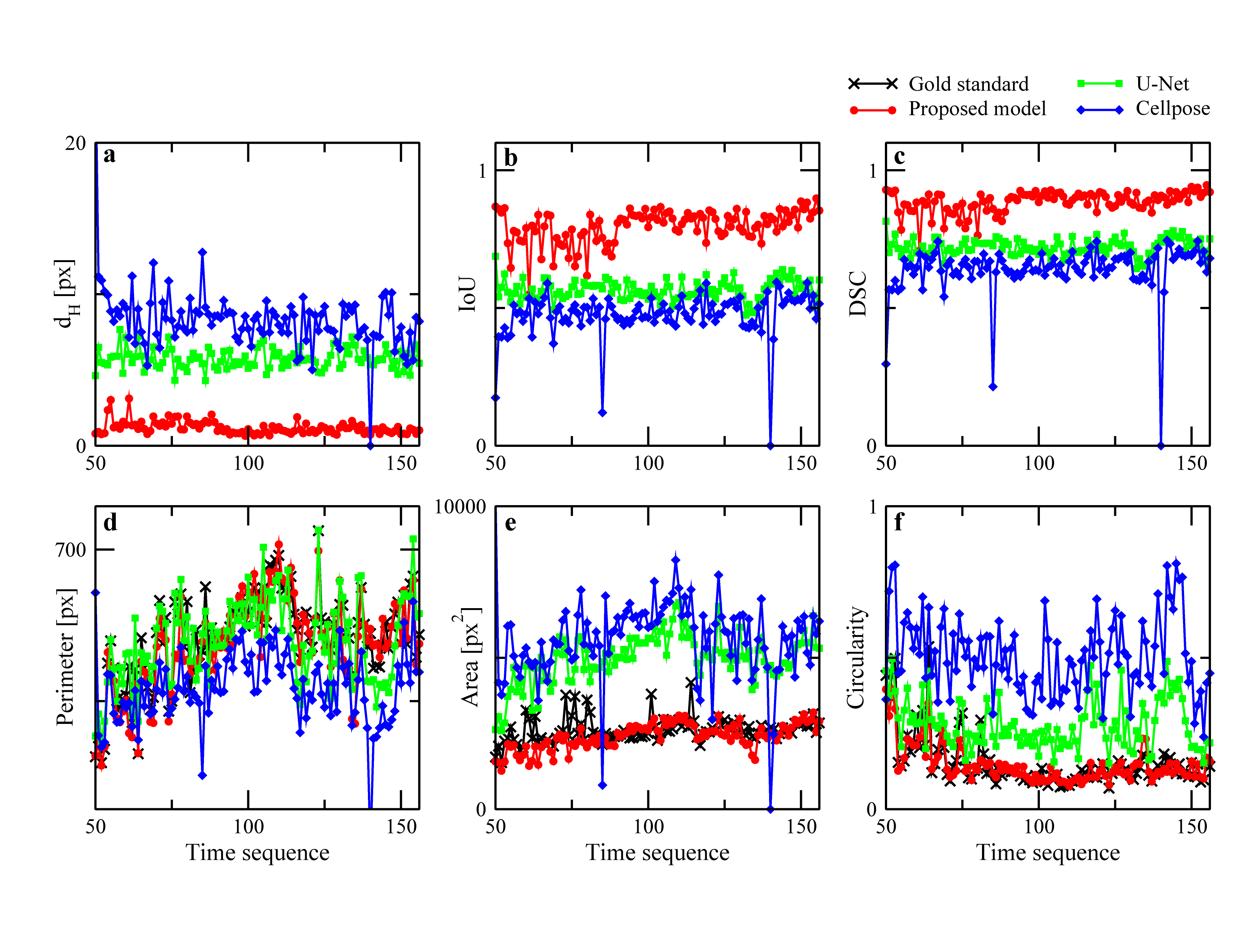}
    \caption{The quantitative comparison for the macrophage denoted by ``ii'' in Fig. \ref{fig_seg_MLcompare}. \textbf{a}: The mean Hausdorff distance from the gold standard, \textbf{b}: IoU (Jaccard) index, \textbf{c}: Sørensen--Dice coefficient, \textbf{d}: Perimeters of segmentations, \textbf{e}: Areas of segmentations, and their \textbf{f}: circularities.  Time sequence in the horizontal axis indicates the order of time frames, and its interval is 2 minutes.}
   \label{fig_seg_quantitative2}
\end{figure}

\indent
The performance of the proposed segmentation method is compared to three different deep learning methods, U-Net \cite{ronneberger2015,falk2019}, Cellpose\cite{stringer2021}, and Splinedist \cite{mandal2021}.
Those three methods have been designed to segment general or flexible shapes in microscopic images.
In their research, they have shown the high performance for different types of biological objects such as biological tissue, microglial cells \cite{falk2019}, elongated dendrites \cite{stringer2021}, and non-star-convex shapes of cell nuclei \cite{mandal2021}.
We choose the training dataset at \url{www.cellpose.org}, which contains not only various biological images but also general objects such as fruits, beans, etc., to train networks for varying shapes of macrophages with minimal manual effort.
We may expect a better performance of the mentioned machine learning methods if they are trained on macrophage data. However, it would need a lot of datasets for training networks to cover the high variability of shapes and image intensity of macrophages. In addition, due to the complex shapes of macrophages, the preparation of training images demands substantial manual effort even though we use the semi-automatic segmentation approach.
Therefore, segmentation using training networks calibrated for macrophages is out of the scope for the present paper.
The toolbox called zerocostdl4mic \cite{von2021democratising} is used to train the networks of deep learning methods, and then the trained networks are applied to the second dataset of this paper. 
Fig. \ref{fig_seg_MLcompare} shows the original and segmented images in two different time slices.
Except for Splinedist, U-Net and Cellpose segment and detect macrophages quite successfully. However, the segmented bodies obtained from U-Net and Cellpose are wider and less accurate than observed in the original images.
The advantage of the wider shapes is that the segmented macrophages can cover the entire shape, as shown in the green rectangles in Fig. \ref{fig_seg_MLcompare}.
However, it can be a problem when two different macrophages are close to each other.
For instance, the two different macrophages inside the red rectangles at $\theta=44$ in Fig. \ref{fig_seg_MLcompare} are segmented properly by our segmentation method, while the results from those two deep learning methods show a connected macrophage. 
The quantitative comparisons are carried out to see details of the difference between the proposed method, U-Net, and Cellpose.
In Fig. \ref{fig_seg_MLcompare}, the macrophage, which has somewhat rounded shapes, is selected and denoted as ``i'' with the pink arrow. In contrast, the macrophage marked ``ii'' has very irregular shapes over time.
The quantitative plots for the ``i'' and ``ii'' macrophages are shown in Fig. \ref{fig_seg_quantitative1} and Fig. \ref{fig_seg_quantitative2} respectively.
For the ``i''  macrophage, the mean Hausdorff distance from the gold standard averaged in time for U-Net, and Cellpose equals $5.29$ and $12.05$, respectively. For the macrophage ``ii'', it is $5.74$ and $8.32$, respectively. In both cases, it is higher than for the proposed method. As expected by Fig. \ref{fig_seg_MLcompare}, the area of U-Net and Cellpose tend to be higher than the gold standard area and the area obtained by our method. In particular, it is more apparent when the shapes of the macrophage are complicated (Fig. \ref{fig_seg_quantitative2}) since U-Net and Cellpose give wider and smoother shapes which yield the high circularity as shown in Fig. \ref{fig_seg_quantitative1} and Fig. \ref{fig_seg_quantitative2}.  
The U-Net method performs better than Cellpose in general; however, it sometimes fails to segment macrophages where Cellpose and our approach can segment them, see blue circles in Fig. \ref{fig_seg_MLcompare}.

\clearpage
\begin{table}[ht]
    \centering
    \renewcommand{\arraystretch}{1.3}
    \begin{tabular}{l ccc ccc}
    \toprule
    & \multicolumn{3}{c}{``i''} & \multicolumn{3}{c}{``ii''} \\
    \cmidrule(lr){2-4} \cmidrule(lr){5-7}
      & $\overline{d_{\text{H}}}$  & $\overline{\text{IoU}}$  & $\overline{\text{DSC}}$  & $\overline{d_\text{H}}$ & $\overline{\text{IoU}}$  & $\overline{\text{DSC}}$  \\
    \midrule
    Proposed  & 1.65 & 0.77 & 0.86  & 1.19  & 0.80  & 0.89 \\
    U-Net & 5.29 & 0.62 & 0.76  & 5.74  & 0.57  & 0.72 \\
    Cellpose  & 12.05 & 0.40 & 0.57  & 8.32  & 0.48 & 0.64 \\
    \bottomrule
    \end{tabular}
    \caption{The average over time of the mean Hausdorff distance $\overline{d_\text{H}}$, the average over time of IoU (Jaccard) index $\overline{\text{IoU}}$, and the average over time of Sørensen--Dice coefficient $\overline{\text{DSC}}$ obtained by using three segmentation methods.}
    \label{table_segment_Hd}
\end{table}
\subsection{Visual and quantitative assessment of tracking}
In this section, we present trajectories of moving macrophages extracted by using our tracking algorithm. We applied the proposed method to two 2D+time datasets described in Introduction. The tracking process for both datasets is applied in the same way: first, macrophages are segmented in every time slice; second, the partial trajectories of cells overlapping in the temporal direction are extracted; last, the partial trajectories are connected using the tangent calculation.

In the first dataset, the macrophages are sparsely distributed in the spatial domain and do not touch each other. Therefore, it is relatively easy to evaluate the tracking performance visually. The partial trajectories are connected using $\Delta r=30$ pixels in (\ref{cond_connect}). 

\begin{figure}[ht]
   \includegraphics[scale=1.2]{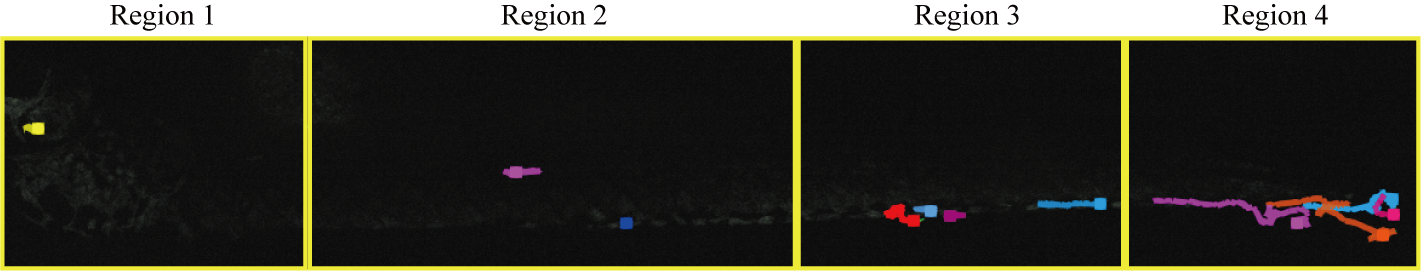}
   \centering
    \caption{Trajectories of macrophage movements at the final time slice $\theta_{F}=74$. The spatial domain is $3755*683$ pixels, and the entire domain is divided into four subdomains. Different colors represent each macrophage and its trajectory.}
   \label{fig_tra_1st_whole}
\end{figure}

\begin{figure}[ht]
   \includegraphics[scale=0.95]{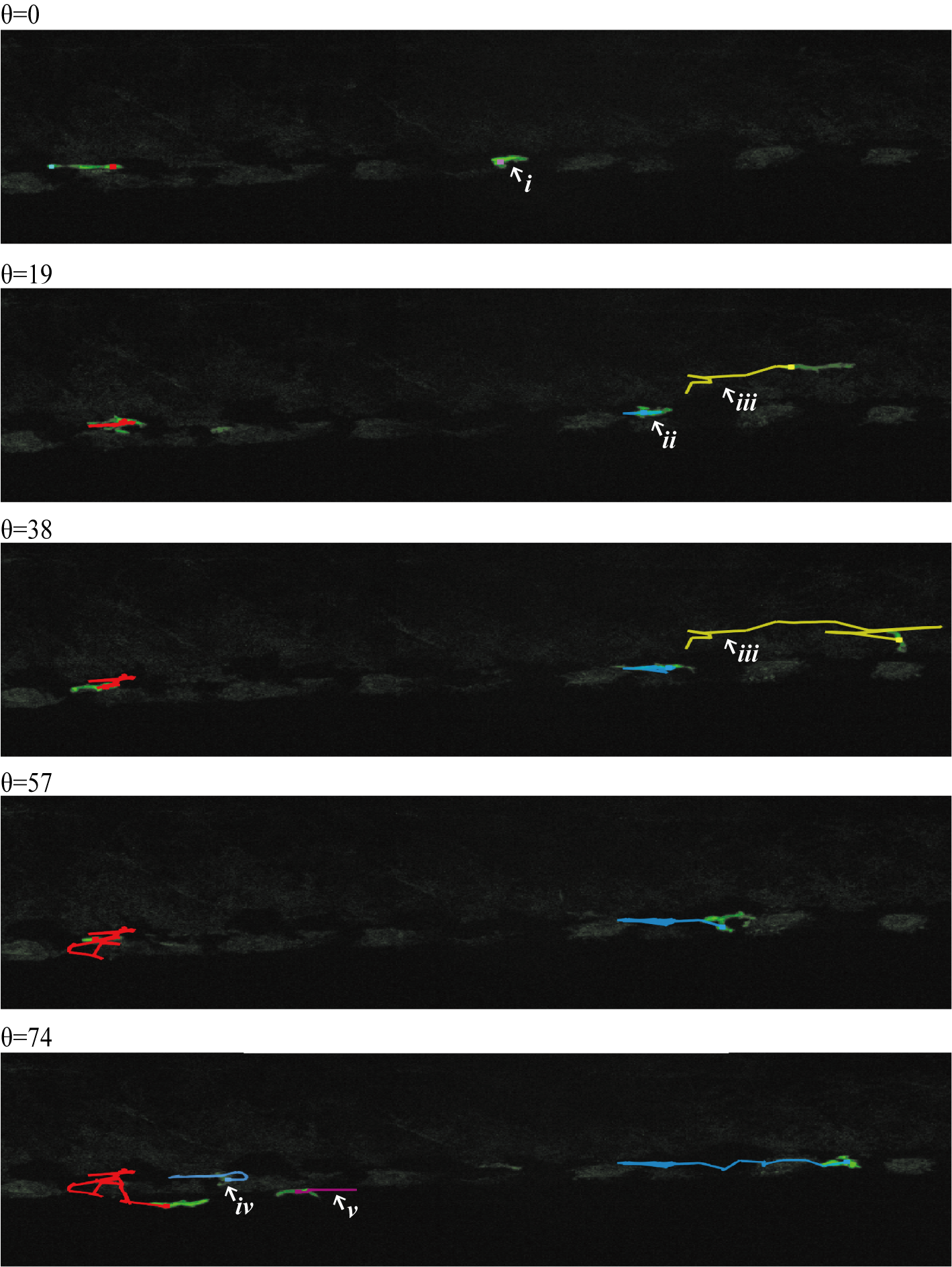}
   \centering
    \caption{Trajectories in subdomain $3$ from Fig. \ref{fig_tra_1st_whole} visualized in five different time slices.}
   \label{fig_tra_1st_R3}
\end{figure}

\begin{figure}[ht]
   \includegraphics[scale=0.95]{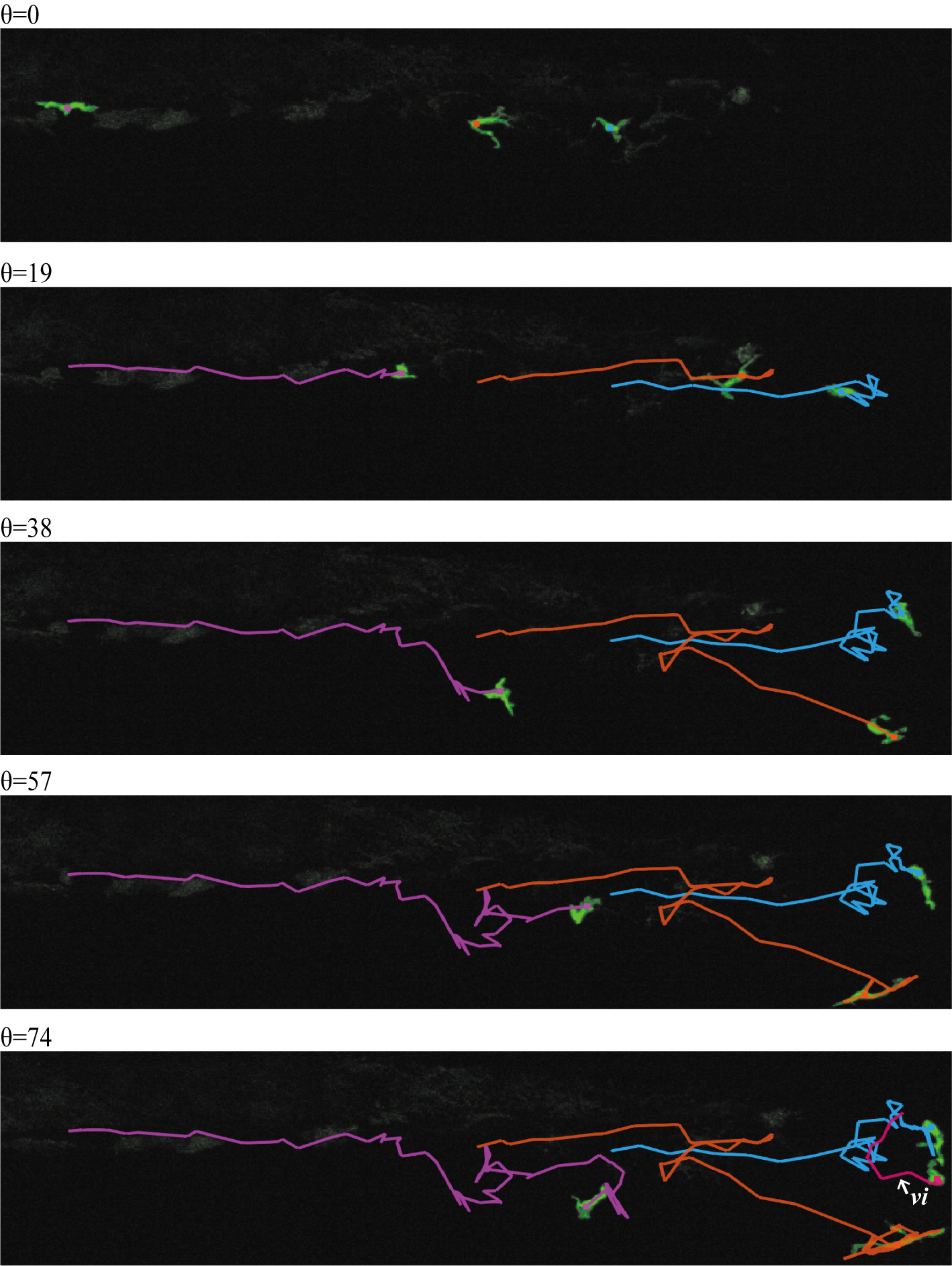}
   \centering
    \caption{Trajectories in subdomain $4$ from Fig. \ref{fig_tra_1st_whole} visualized in five different time slices.}
   \label{fig_tra_1st_R4}
\end{figure}

\begin{figure}[ht]
   \includegraphics[scale=0.93]{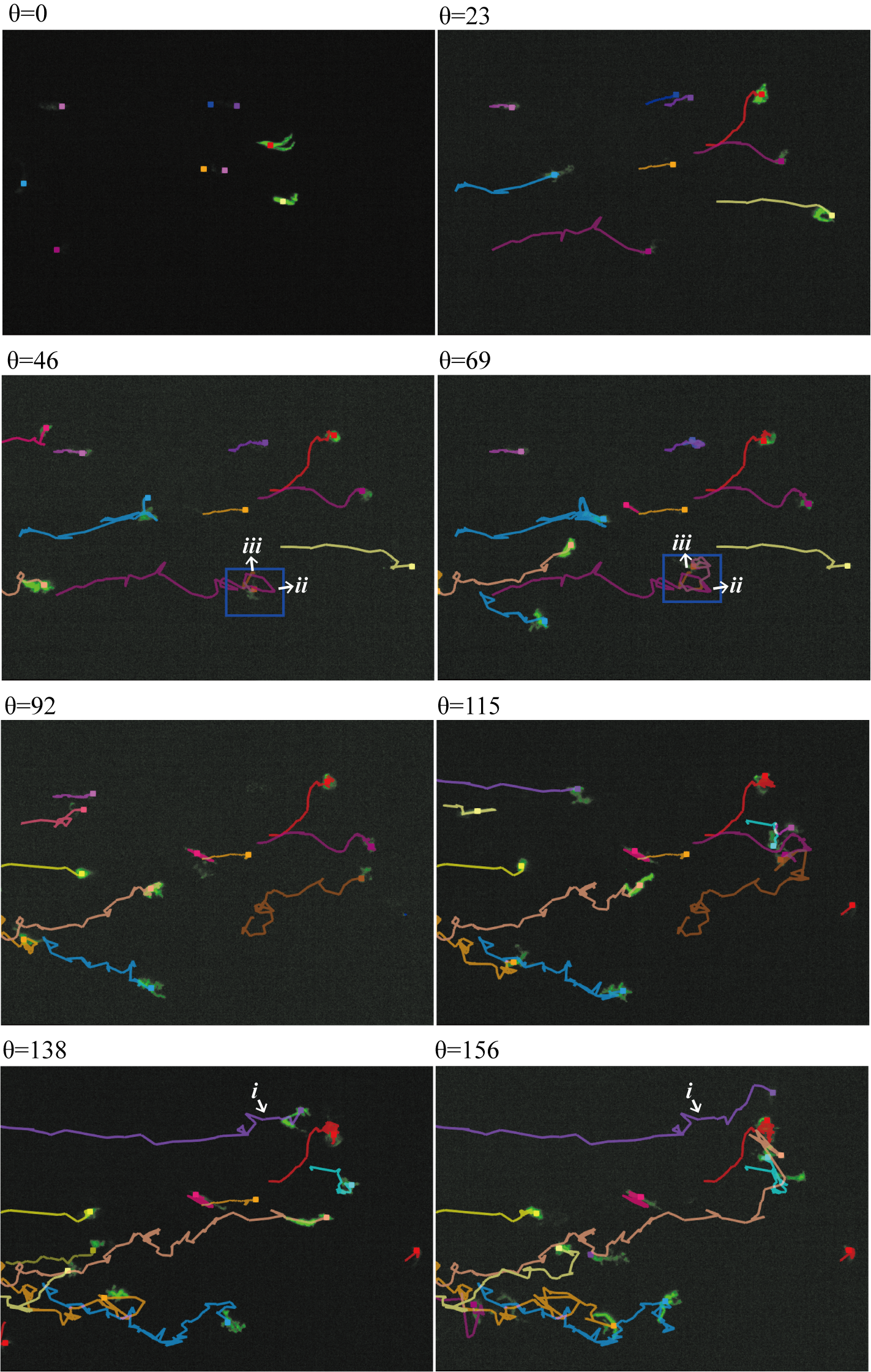}
   \centering
    \caption{Final trajectories in eight subsequent time slices. The final time slice of the second dataset is $\theta_{F}=156$. Here, the size of the whole spatial domain is $1758*1306$ pixels.}
   \label{fig_tra_2nd}
\end{figure}

Fig. \ref{fig_tra_1st_whole} shows trajectories obtained by our tracking algorithm visualized at the final time slice in 4 chosen subdomains of the 2D image.
In Figs. \ref{fig_tra_1st_whole}--\ref{fig_tra_2nd}, square dots denote the position of approximate cell centers in the visualized time slice, and lines show the macrophage trajectories from its appearance up to the visualized time slice. 
The macrophages move more actively as they locate on the right side since the site of the wound is on the rightmost. 
In Regions $1$ and $2$, the macrophages migrate in the way of ``random movement'', whereas most macrophages show directional motion in Regions $3$ and $4$.
Fig. \ref{fig_tra_1st_R3} and Fig. \ref{fig_tra_1st_R4} are chosen to see the trajectories of macrophages in detail. 
In Fig. \ref{fig_tra_1st_R3}, there are three detected approximate cell centers at $\theta=0$ and the left-most one disappears after $4$ time slices. The macrophage denoted by light purple (also denoted by ``\textit{i}'') changes to the one marked in sky blue (``\textit{ii}'') as shown in the second panel. These two partial trajectories (``\textit{i}'' and ``\textit{ii}'') are not connected since they move very fast, i.e., the distance between the estimated points from the two partial trajectories is too far.
The macrophage denoted by yellow (``\textit{iii}'') moves actively compared to other macrophages in Region 3 as shown in the second and third panel. In the last panel of Fig. \ref{fig_tra_1st_R3}, two macrophages (``\textit{iv}'' and ``\textit{v}'') appear after $\theta=57$ and keep showing until $\theta=74$.
The trajectories in Region 4 are visualized in Fig. \ref{fig_tra_1st_R4} from the beginning to the final time slice ($\theta=74$) and show partially ``random movement'', but they migrate dominantly toward the site of the wound. It shows that the tracking algorithm can cover both situations of random movement and directional movement. 
One macrophage denoted by the pink line (``\textit{vi}'') appears after $\theta=57$ in the upper right and moves opposite direction from the wound a bit, but it comes back to the right side.

\begin{figure}[ht]
   \includegraphics[scale=0.85]{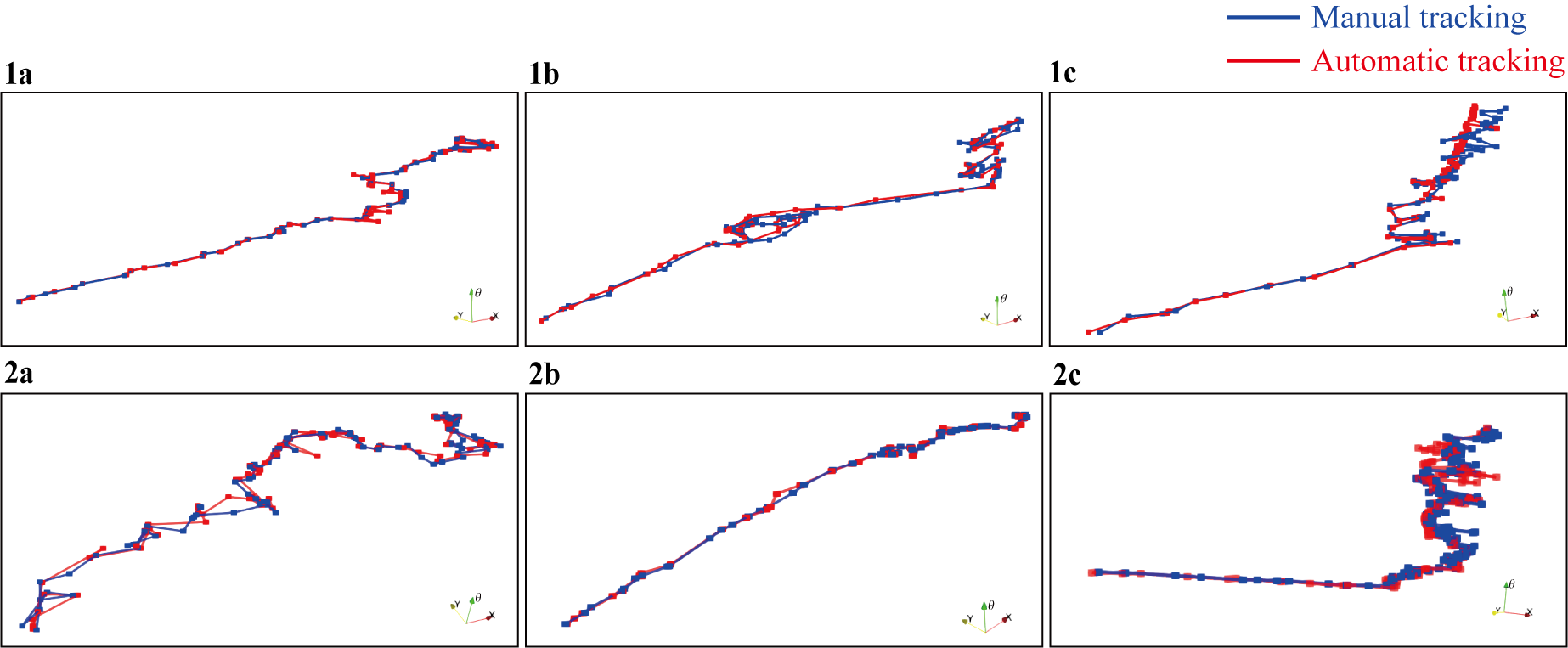}
   \centering
    \caption{Extracted trajectories of three different macrophages from manual and automatic tracking methods in the first \textbf{1a}--\textbf{1c} and second dataset \textbf{2a}--\textbf{2c}. The number of points of \textbf{1a}--\textbf{1c} is $75$ and the numbers of points of \textbf{2a}--\textbf{2c} are $109$, $57$, and $157$, respectively. The blue lines show manual tracking in Fiji \cite{schindelin2012fiji}, and the red lines show our proposed tracking method.}
   \label{fig_tra_HD}
\end{figure}

The overall number of partial trajectories was $165$, and it decreased to $32$ after connection using tangent approximation. 
To quantify the accuracy of tracking, the mean Hausdorff distance between trajectories obtained from our automatic tracking algorithm and the manual tracking performed by Fiji software \cite{schindelin2012fiji} was computed. Three different macrophages of the first dataset, having clear signal over the whole time sequence, were selected for comparison and results are presented in Fig. \ref{fig_tra_HD} left column. The mean Hausdorff distances of the automatic and manual trajectories presented in panels \textbf{1a}--\textbf{1c} of Fig. \ref{fig_tra_HD}  were $1.17$, $1.55$, and $1.19$ pixels, respectively. Also, the average distance between the points at each time slice of two trajectories obtained from the manual and the proposed method are computed (Table \ref{table_track_Hd}).  These distances are very small compared to the overall length of trajectories, indicating the high accuracy of the automatic tracking algorithm for the first dataset. 

\clearpage
The second dataset is much more noisy, and the macrophage movement is more complicated. In case the image intensity of macrophages is very weak or has high variability, the segmentation yields several fractions of a single macrophage, see Fig. \ref{fig_seg3}.
The segmented fractions cause the existence of more than one partial trajectory in the same macrophage in the several common time slices, as illustrated in Fig. \ref{fig_track_connection}\textbf{c}. For this dataset, the condition in Equation \ref{cond_connect} with $\Delta r = 60$ was first used to connect the partial trajectories and then parameters $\Delta r_{2} = 120$ and $\Delta r_{\theta}=5$ were used to avoid closed trajectories due to macrophage segmentation split. 
We note that $\Delta r_{2}$ can be chosen less sensitively than $\Delta r$ since many trajectories are already connected.

The final trajectories of the second dataset in the whole spatial domain are visualized in Fig. \ref{fig_tra_2nd}.
There are $9$ detected macrophages at the beginning, and $12$ macrophages are shown at the last time slice as new macrophages appear and disappear over time.
The site of the wound is positioned on the right side, and many macrophages migrate toward the wound. Especially, the macrophage denoted by violet (``\textit{i}'') at the top of the pictures from $\theta=115$ to $\theta=156$ shows very fast movement, which implies the macrophage yields many partial trajectories, and our method enables us to connect them.

In the second row of the figure, the two trajectories (purple, ``\textit{ii}'' and brown, ``\textit{iii}'') inside the blue rectangle are located in the same cell. However, they are shown differently because the segmentation cannot extract the entire shapes of macrophages, and their number of the common time slices $\Theta_{c}$ is greater than $6$.

The number of trajectories before and after the first connection by using condition in Equation \ref{cond_connect} (see also Fig. \ref{fig_track_connection}\textbf{a}--\textbf{b}) was $930$ and $234$, respectively. After the second connection of closed common trajectories (see Fig. \ref{fig_track_connection}\textbf{c}--\textbf{d}),  the number of extracted trajectories decreased again significantly to $69$.
In addition, the average length of extracted trajectories after the second connection increased from $69.53$ to $363.29$ in pixel units.
 
The mean Hausdorff distance computed for three selected trajectories of the second dataset, see panels \textbf{2a}--\textbf{2c} of Fig. \ref{fig_tra_HD}, was $3.13$, $4,35$, and $2.40$ in units of pixels, which is very low in comparison to the average length of trajectories. It shows the high accuracy of tracking for this dataset again. The comparison of the average distance computed by averaging the Euclidean distance between points at each time frame of manual and the proposed tracking is presented in Table \ref{table_track_Hd}. This distance is bigger than the mean Hausdorff distance but still small compared to the average size of macrophages.

\begin{figure}[ht]
   \includegraphics[scale=0.85]{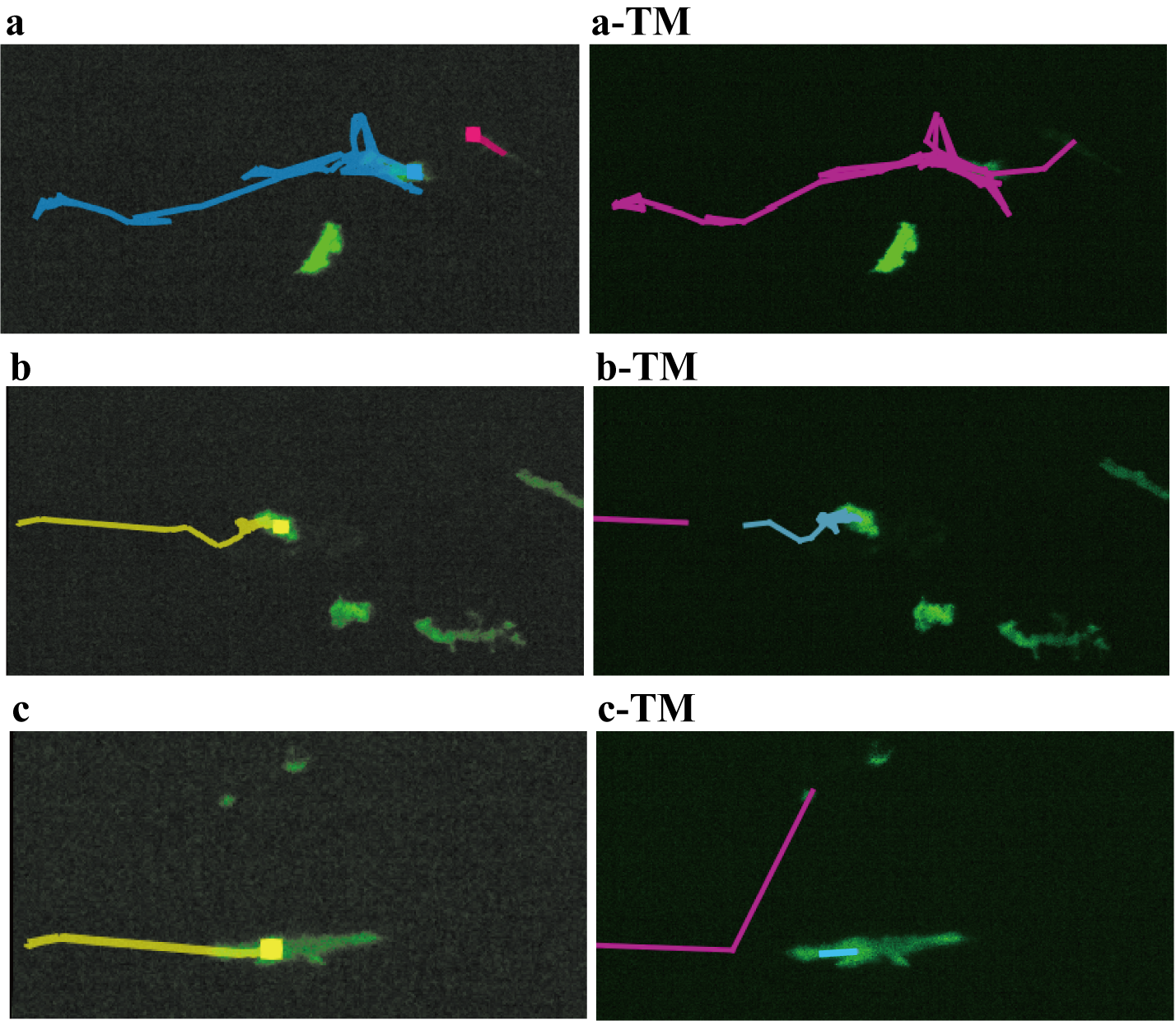}
   \centering
    \caption{The trajectories obtained from the proposed method (left column) and the LAP method in TrackMate in Fiji \cite{tinevez2017} (right column). The panels of \textbf{a},\textbf{b},\textbf{c} and  \textbf{a-TM},\textbf{b-TM},\textbf{c-TM} show trajectories extracted at different time moments, $\theta=69$, $\theta=156$, and $\theta=90$, respectively. }
   \label{fig_TM_track}
\end{figure}

\begin{figure}[ht]
   \includegraphics[scale=0.85]{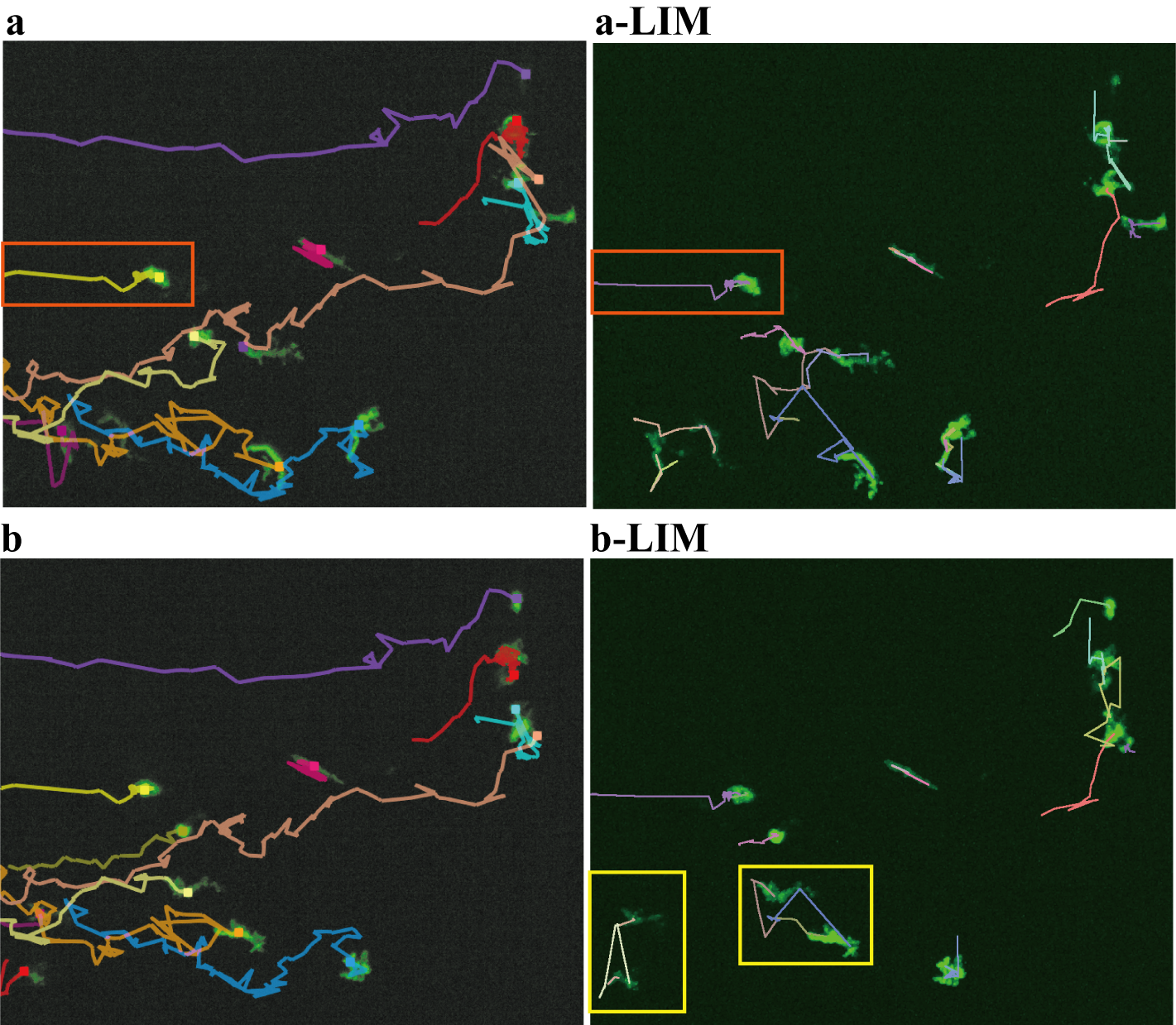}
   \centering
    \caption{The trajectories obtained from the proposed method (left column) and the LIM tracker \cite{aragaki2022} (right column). The panels of \textbf{a}, \textbf{b} and \textbf{a-LIM}, \textbf{b-LIM} show different time moments at $\theta=156$, $\theta=147$, respectively. }
   \label{fig_LIM_track}
\end{figure}

Further check of automatic tracking accuracy we performed by counting the number of correct and wrong links in every time slice by visual inspection.
If a trajectory is linked correctly to the same macrophage in the next time slice, we count it as a correct link. However, we consider the wrong link for a trajectory when it is linked to a different macrophage or it disappears in the next time slice although segmented regions exist for the corresponding macrophage. 
We define the time slice tracking accuracy as the ratio between the number of correct links and the total number of links detected visually in one time slice in a forward manner, and we define the mean accuracy of tracking as the average of the tracking accuracy over all time slices.
As a result, the mean accuracy of tracking in the first dataset was $0.975$ and in the second dataset, it was $0.974$. Both results demonstrate that the suggested tracking method is able to achieve high accuracy bigger or equal to $97.4\%$ for generic datasets obtained by confocal microscopy. 
The final results of cell tracking are also provided as two videos in the supplementary materials (\url{ https://doi.org/10.1016/j.compbiomed.2022.106499}).
The video named 1st\_dataset.mov shows the trajectories in Region 4 in the first dataset.  
For the second dataset, 2nd\_dataset.mov shows the obtained trajectories in the whole spatial domain.
\newline
\indent
The trajectories obtained by the proposed tracking method are compared to the results of TrackMate \cite{tinevez2017} and LIM Tracker \cite{aragaki2022}, which can be easily implemented in Fiji.
For the cell detection in TrackMate, we used the LoG detector with the parameter values for ``Estimated object diameter'' = $110$ and ``Quality threshold'' = $0.004$. Then, the LAP tracker was applied for ``linking with MAX distance'' equals to $200$ pixels and ``Gap closing'' = $100$ pixels with ``Max frame gap'' = $2$.
The other tracking algorithms, the Kalman tracker and Nearest-neighbor tracker, in the Fiji plugin were also applied. However, the LAP tracker gave the least number of disconnected trajectories so we will discuss only the LAP tracker results.
The mean accuracy of the LAP tracker in Fiji was evaluated in the same manner as measuring in the proposed method. It yielded 0.971, indicating the majority of the trajectories obtained from the LAP tracker correctly represented the movement of macrophages, similar to the proposed method. However, we point out the cases where this method failed to link macrophages correctly, but the proposed method was successful. 
The majority of the trajectories obtained from the LAP tracker correctly represented the movement of macrophages, similar to the proposed method. However, we point out the cases where this method failed to link macrophages correctly, but the proposed method was successful.
In Fig. \ref{fig_TM_track}, the visualizations of trajectories obtained by our tracking method at different time slices are shown in the left column (denoted by \textbf{a}--\textbf{c}), and the results obtained by LAP tracker are shown in the right column (denoted by \textbf{a-TM}--\textbf{c-TM}).
Fig. \ref{fig_TM_track}\textbf{a} shows two different trajectories where one macrophage moves actively (blue curve) and the other moves very slowly (pink curve). These two macrophages are located close to each other in the previous time slice. Therefore LAP links them due to their close distance, as shown in the panel \textbf{a-TM}. On the other hand, the proposed tracking method links the points correctly since there are overlapping segmented shapes for each trajectory.
The panels denoted by \textbf{b} and \textbf{b-TM} in Fig. \ref{fig_TM_track} show the case when the LAP tracker fails to link the two disconnected trajectories (see purple and blue curves in panel \textbf{b-TM}).
The distance between the endpoints of those two trajectories in \textbf{b-TM} is considerable because the macrophage moves fast at the time slice in which the disconnection occurs. 
In the proposed method, the trajectory is successfully connected since the algorithm considers the direction of movement when it links partial trajectories.
To check if those two trajectories can be connected by the LAP method, we slightly increased the linking parameter ``linking with MAX distance'' to $202$. 
The increase of the parameter does not help to connect the trajectories; moreover, it causes another wrong connection at time slice $\theta=90$ as shown in the panel of \textbf{c-TM}. 
\newline
\indent
In addition, the comparison between the proposed tracking method and the LIM tracker \cite{aragaki2022} implemented in Fiji is shown in Fig. \ref{fig_LIM_track}. 
We used the automatic cell detection provided by the software setting the parameters to ``Threshold'' = $14000$, ``Cell size'' = $50$, and ``ROI size'' = $110$. Then, for the automatic cell tracking, the parameters ``Link ROI range'', ``Link split track'', and ``Fill frame gap'' were set to $200$, $70$, and $2$, respectively. 
The panels denoted by \textbf{a} and \textbf{a-LIM} show the results obtained by our method and the LIM Tracker at the final time slice, respectively.
The trajectory inside the orange rectangle is very similar to the result of our tracking method, unlike the one obtained from the LAP tracker, cf. Fig. \ref{fig_TM_track}\textbf{b}. However, many trajectories are disconnected since the cell size in this software can be set maximally only to $50$ pixels which are not sufficient for our datasets.
Therefore, the trajectories obtained from the LIM tracker gave a lower mean accuracy, 0.893.
The panels of \textbf{b} and \textbf{b-LIM} show trajectories at $\theta=147$ obtained by our method and the LIM tracker, respectively.
Similarly to the case of \textbf{c-TM} in Fig.\ref{fig_TM_track}, there are several wrong connections shown inside the yellow rectangles in \textbf{b-LIM} panel.

\clearpage

\begin{table}[ht]
    \centering
    \renewcommand{\arraystretch}{1.3}
    \begin{tabular}{c || c | c | c | c | c }
    \toprule
      &  \# of points & $d_{\text{H}}$ [px] & $ d_{\text{avg}}$ [px] & $L_{\text{manual}}$ [px] & $L_{\text{auto}}$ [px] \\  \hline
      \textbf{1a} & 75 & 1.17 & 4.00 & 653.21 & 842.97 \\  \hline
      \textbf{1b} & 75 & 1.55 & 7.45 & 993.90 & 848.92 \\ \hline
      \textbf{1c} & 75 & 1.19 & 6.20 & 765.59 & 570.82 \\  \hline
      \textbf{2a} & 109 & 3.13 & 17.33 & 2425.88 & 3026.21 \\  \hline
      \textbf{2b} & 57 & 4.35 & 19.78 & 1887.99 & 1920.75 \\  \hline
      \textbf{2c} & 157 & 2.40 & 15.71 & 2039.42 & 1720.62 \\ 
    \bottomrule
    \end{tabular}
    \caption{The two different types of the distance between trajectories obtained from manual and proposed tracking. $d_{\text{H}}$ and $d_{\text{avg}}$ denote the mean Hausdorff distance between two curves and the average of the Euclidean distance between two points at each time slice. $L_{\text{manual}}$ is the total length of trajectories by manual tracking, and $L_{\text{auto}}$ is the total length obtained from proposed tracking.}
    \label{table_track_Hd}
\end{table}

\clearpage
\section{Discussion}
In this paper, we presented a new approach to cell tracking based on image segmentation. The proposed segmentation and tracking method was performed in 2D + time microscopy data.
\newline
\indent
Segmentation is carried out in three steps: space-time filtering, the local Otsu's thresholding, and the SUBSURF approach. 
The dataset we dealt with has different intensities of the background noise and macrophages by time slices. Since the image intensity of some macrophages is very close to the background noise, the first task was to make distinguishable the signals between the background noise and macrophages through space-time filtering. Then, the second task was capturing the proper shapes of macrophages having huge variability of image intensity. To do this, we applied the thresholding technique locally by using the local Otsu's method. Lastly, the SUBSURF method eliminates the artifacts that occur after applying the local Otsu's method.
As a result, we showed that the proposed segmentation enables us to segment all macrophages.
Moreover, the proposed segmentation was compared to machine learning-based segmentation, U-Net, and Cellpose and different local threshold methods (see Supplementary materials \url{ https://doi.org/10.1016/j.compbiomed.2022.106499}).
The comparison was presented using the mean Hausdorff distance, IoU index, and Sørensen--Dice coefficient between the results of different segmentation methods and the gold standard for two macrophages. 
The three measurements showed the proposed method gives a reasonable performance, especially for complex shapes of the macrophage. 
\newline
\indent
Based on the segmented images, we performed tracking with the proposed method in two different datasets. 
The macrophages often do not overlap continuously in the temporal direction, therefore we traced them in two stages.

The partial trajectories were extracted first by checking segmented regions of macrophages are overlapped in the temporal direction.
Then, the direction of the movement, by computing the tangent of the partial trajectories, was approximated. 
We assumed the macrophages tend to keep their direction of movement at the time point when their segmented bodies are disconnected in the temporal direction.
Hence, we connected partial trajectories when they have a similar direction of movement and they are close to each other.

The performance of tracking was analyzed by comparing trajectories obtained from manual and automatic tracking in three ways; computation of the mean Hausdorff distance, the average distance at each time slice, and the mean accuracy.
The mean Hausdorff distance and the average of the Euclidean distance showed small differences from the trajectories of manual tracking compared to the total length of trajectories and size of macrophages.
The mean accuracy was defined by the average ratio between the correct and total links at each time slice. It showed high accuracy of  $97.5$\% in the first dataset and $97.4$\% in the second dataset from the proposed tracking. 
We also compared with other tracking methods, TrackMate and LIM tracker, and presented the cases when the proposed method gave more accurate results.
\subsection*{Limitations}
There are still open questions about the proposed method. First,  the proposed segmentation method sometimes fails to extract an entire body of a macrophage. It mainly occurs when macrophages stretch their body a lot, causing weak intensity inside the macrophages. In this situation, the local window centered by the part of the weak intensity determines that it does not contain macrophages. SUBSURF does not connect these segmented fractions corresponding to the same macrophage since the fractions are quite far apart. 
\newline
\indent
Second, the performance of tracking is dropped with poor results of the segmentation. It is quite apparent that segmentation determines the robustness of segmentation-based tracking. In our case, the segmentation problem was mainly due to segmented fractions for a single macrophage. To solve it, we considered the number of time slices appearing simultaneously for two close trajectories and linked them as illustrated in Fig. \ref{fig_track_connection}. However, there is a case when the two partial trajectories fail to be connected, as presented in the blue square in Fig. \ref{fig_tra_2nd} because of the large number of common time slices. This situation can happen if several fractions are more than two, yielding many partial trajectories. Therefore, the proposed tracking still relies on the segmentation quality even though we tried to deal with the segmentation problem. 
\section{Conclusion}

We proposed automated methods for the segmentation and tracking of macrophages with highly complex shapes and migration patterns.
We described the proposed methodology, presented the results, and discussed the performances and limitations.
The method could be improved by considering the points described in the section Limitations. Also, we suggest the possible applications of the proposed method. 
\subsection*{Future works and possible applications}
In order to segment the entire shapes of macrophages, the global information of individual macrophages should also be considered together with considering local information from the proposed method. The space-time segmentation by adding the time component in the segmentation \cite{uba20224d} could improve the performance. 
For tracking, the best is to segment macrophages as accurately as possible, but it is also necessary to think of how to obtain more information from the partial trajectories to overcome the low quality of the segmentation.
\newline
\indent
By segmenting and tracking automatically, we expect that the proposed method can provide quantitative data and evidence to figure out how relevant the polarization modes of macrophages are to shapes and patterns of migrations \cite{sipka2022macrophages}.
Also, we expect that the proposed tracking model can be applied to macrophages in other animal models and neutrophils in fluorescent images. However, it would need parameter optimization for different datasets.

\section*{Acknowledgement}
\noindent This work has received funding from the European Union's Horizon 2020 research and innovation programme under the Marie Sk\l{}odowska-Curie grant agreement No 721537 and by the grants APVV-19-0460, VEGA 1/0436/20.

\appendix 
\begin{appendices} \label{appendix}

\counterwithin{figure}{section}
\counterwithin{equation}{section}
\section{Parameter optimization in image segmentation}
Among the parameters required for the image segmentation, we chose the eight playing a major role in the segmentation, namely $\tau_{F}$, $K$, $\sigma$ in space-time filtering, $s$, $\delta$ in the local Otsu's, and $\tau_{S}$, $K$, $\sigma$ in the SUBSURF methods. Each parameter has the four chosen values (see Fig. \ref{appendix_fig3}) and it leads to the number of $65536$ combinations in total. The four types of images were selected to measure the performance of segmentation in various circumstances. The three macrophages were selected as having different properties, such as rounded shape, complex shape, and weak image intensity. Also, a part of the background that does not appear any macrophage over time was cropped to check additionally whether the background noise is segmented as an object or not.   
The accuracy of the automatic segmentation was computed by comparing it with the images obtained by the semi-automatic segmentation (gold standard) \cite{mikula2018}.  

\begin{figure}[ht]
    \centering
    \includegraphics[scale=0.5]{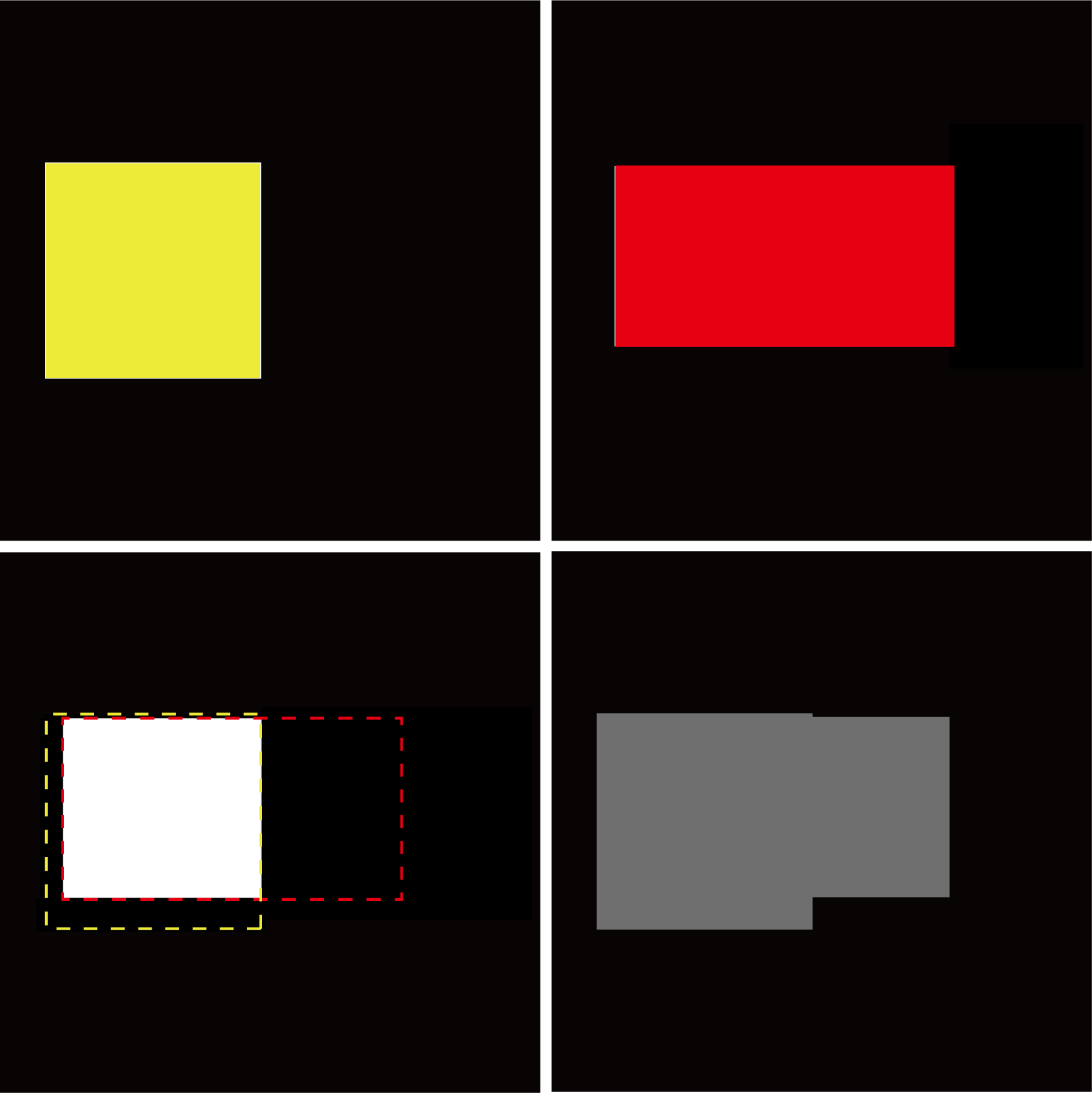}
    \caption{First row: two different objects colored by yellow and red. Second row: the region of intersection and union are shown in white and gray, respectively.  }
    \label{appendix_fig0}
\end{figure}

The accuracy of segmentation was evaluated by measuring the Intersection over Union (IoU), also known as the Jaccard index \cite{jaccard1912distribution}, defined as \textit{Area of intersection}/\textit{Area of union} between the objects given by the gold standard and the proposed segmentation method.
For instance, Fig. \ref{appendix_fig0} shows two objects plotted in yellow and red in the first row. The intersection and union of the two objects are shown in white and gray pixels in the second row of the figure. Therefore, IoU for the two rectangles can be calculated by \textit{number of white pixels}/\textit{number of white and gray pixels}, and it gives $0.5$ in this illustrative example.

Likewise, IoU was computed for four different types of segmented images. Let us denote by $\Phi_k$, $k=1,2,3,4$, the four types of mentioned images, and let $i$ be the index of the time frame.
Then, IoU for $k^{th}$ type of image at $i^{th}$ time frame can be written as 
\begin{equation} \label{appendix_IoU}
    IoU(\Phi_{k},i)=A(I,i)/A(U,i),
\end{equation}
where $A(I,i)$ and $A(U,i)$ represent the number of pixels (area) of the intersection and the union, respectively. 
Next, we define the score of the segmentation for the four types of images by averaging IoU over time. The score for three macrophages is given by 
\begin{equation} \label{appendix_score_mp}
  S(\Phi_{k})=\frac{1}{M_{k}}\sum_{i=1}^{M_{k}}IoU(\Phi_{k},i), \quad k=1,2,3.  
\end{equation}

For the images of the background, we define the score differently since the gold standard gives ``empty'' images without any object. In the background images, $A(I,i)$ is defined by the number of pixels inside the segmented regions, and $A(U,i)$ is given by the number of pixels in the whole image domain. 
Since the score should be decreased when the background noise is segmented as an object, we define it by 
\begin{equation} \label{appendix_score_bkgd}
S(\Phi_{k})=\frac{1}{M_{k}}\sum_{i=1}^{M_{k}}(1-IoU(\Phi_{k},i)), \quad k=4.    
\end{equation}

Here, the number of time steps for each of four images  are $M_{1}=120$, $M_{2}=99$, $M_{3}=61$, and $M_{4}=53$ .
Finally, we measure the mean accuracy $M$ by averaging the scores for the four types of images in every combination of parameters such that 
\begin{equation} \label{appendix_meanAcc}
M = \sum_{k=1}^{4}S(\Phi_{k})/4.    
\end{equation}

\begin{figure}[ht]
    \centering
    \includegraphics[scale=0.65]{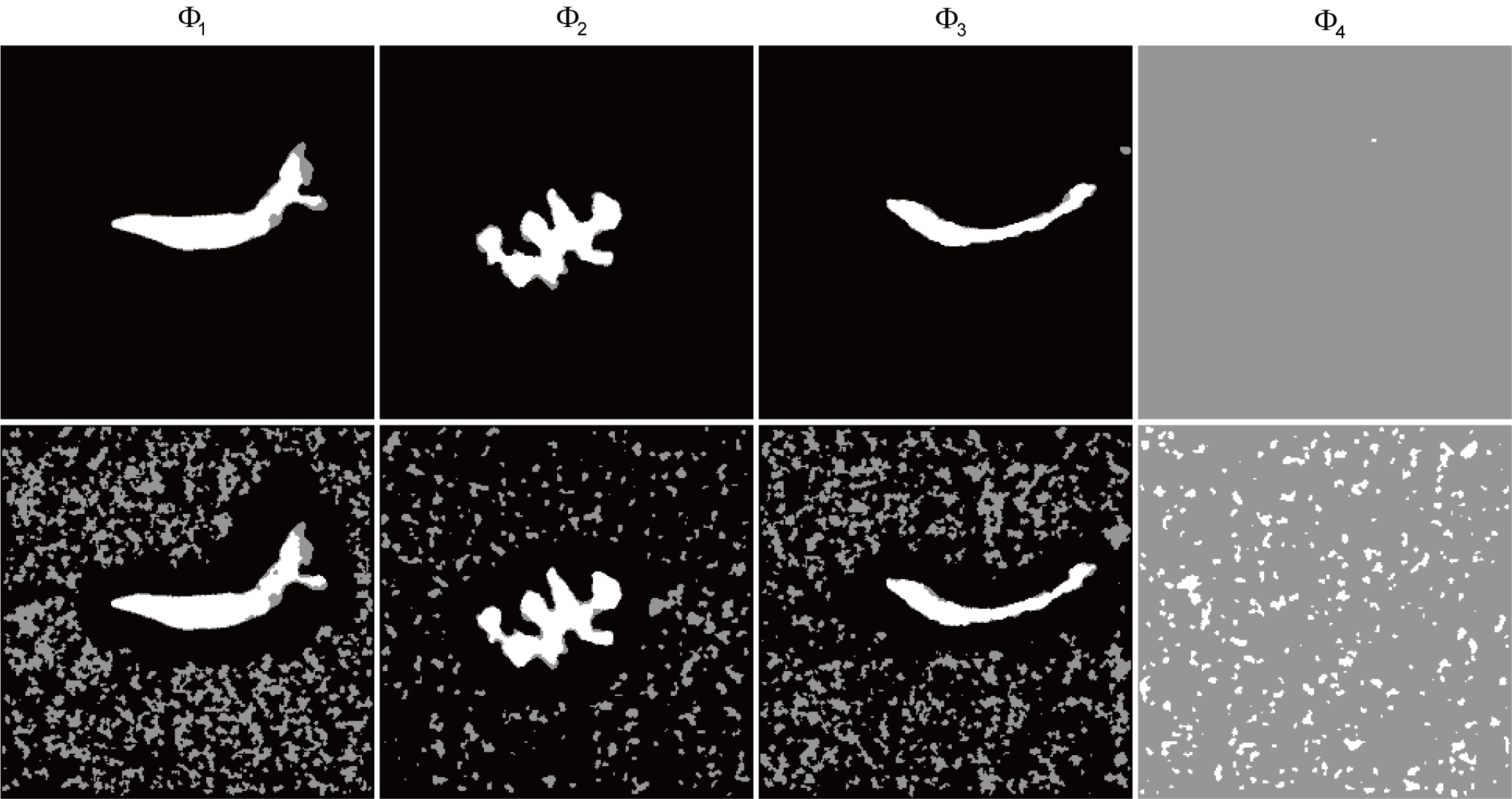}
    \caption{The segmented images of three different macrophages $\Phi_{1}$, $\Phi_{2}$, $\Phi_{3}$ and the background $\Phi_{4}$ by using two combinations of parameters. The mean accuracy of the combination in the top and bottom row is $0.81$ and $0.35$, respectively. }
    \label{appendix_fig1}
\end{figure}

Fig. \ref{appendix_fig1} shows overlapped images obtained by the gold standard and the proposed method with two different combinations of parameters, one optimal and one with very low accuracy. The parameters are described as follows. For space-time filtering; left column: $\tau_{F}=0.25$, $K=100$, $\sigma=0.1$, right column: $\tau_{F}=1$, $K=1000$, $\sigma=0.1$.
For the local Otsu's method; left column: $s=50$, $\delta=0.5$, right column: $s=30$, $\delta=0.3$. 
For the SUBSURF method; left column: $\tau_{S}=0.25$, $K=10$, $\sigma=1$, right column: $\tau_{S}=0.25$, $K=2000$, $\sigma=1$.
In the figure, $A(I,i)$ and $A(U,i)$ are shown by white and gray pixels. 
The mean accuracy $M$ in the top row of Fig. \ref{appendix_fig1} is about $0.81$. It shows that the segmented macrophages from the proposed method are similar to those of the gold standard, and almost no background noise is segmented around the macrophages. 
In the panel of $\Phi_{4}$, the size of segmented background noise is only $6$ pixels. Whereas segmented results in the bottom row, where the mean accuracy is only about $0.35$, show that the background noise is segmented in all panels meaning that this combination of parameters with the low mean accuracy is not able to segment macrophages solely.

\begin{figure}[ht]
    \centering
    \includegraphics[scale=0.7]{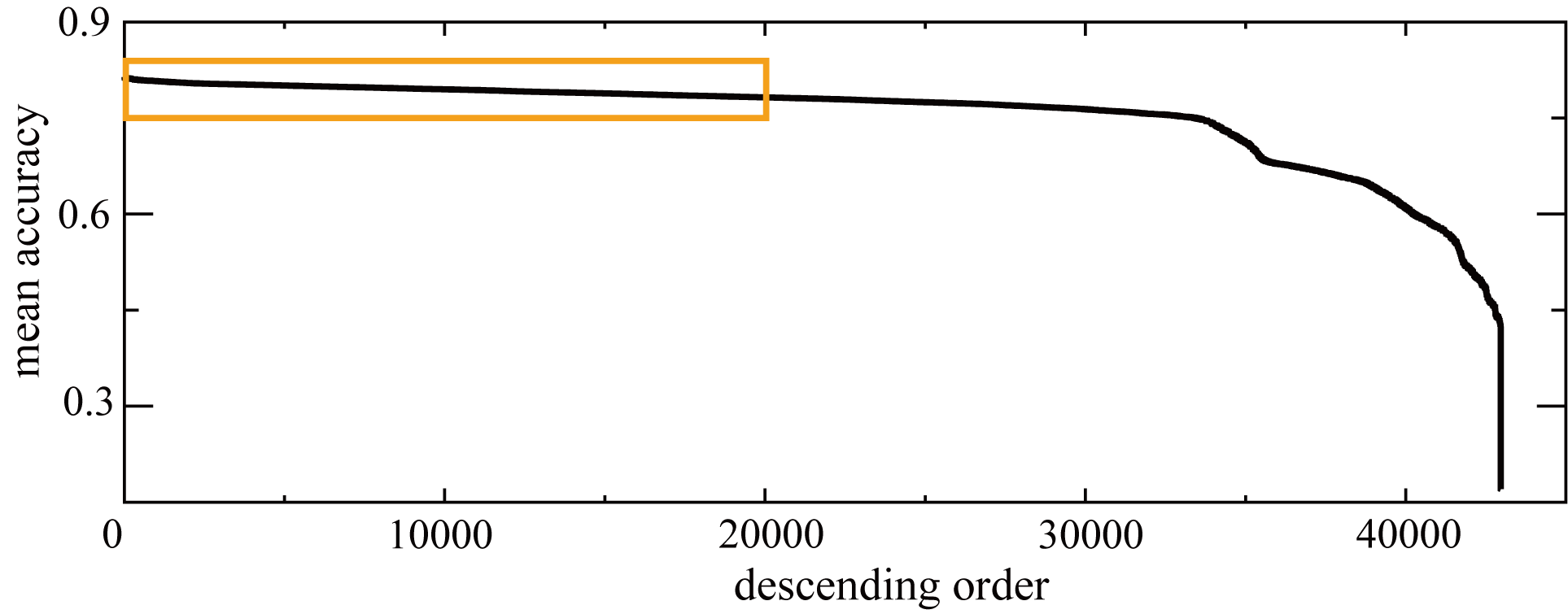}
    \caption{The mean accuracy is sorted from highest to lowest when the score is greater than $0.15$ in three or more consecutive time frames. }
    \label{appendix_fig2}
\end{figure}

\begin{figure}[ht]
    \centering
    \includegraphics[scale=0.8]{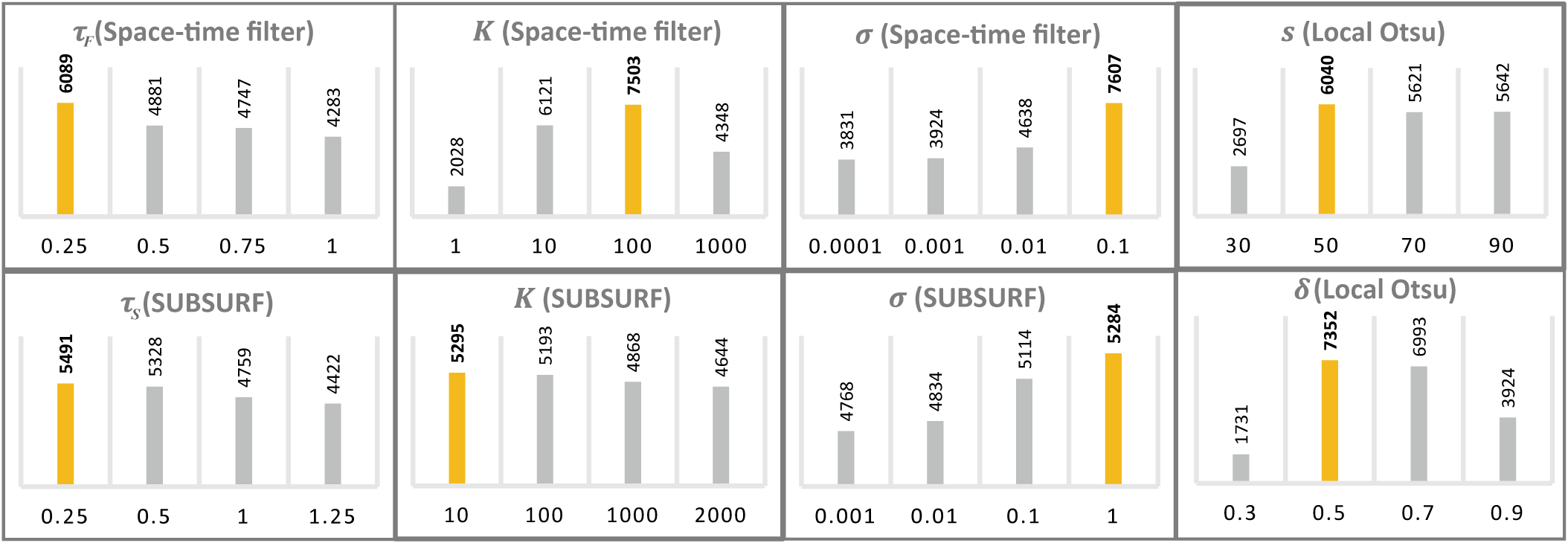}
    \caption{The frequency of each parameter within a range of the orange rectangle in Fig. \ref{appendix_fig2}. The most frequently appearing parameters are denoted by the yellow bars, and they are chosen as optimal. }
    \label{appendix_fig3}
\end{figure}

We first excluded the combinations for which $IoU(\Phi_{k},i)$ is lower than a certain threshold in three or more consecutive time frames to select the optimal parameters. 
There can be a situation where $S(\Phi_{k})$ is high enough, but the value of $IoU(\Phi_{k},i)$ is extremely low in a few time frames, meaning the area of segmented macrophages is very small in those time frame.
We should avoid this situation since segmentation results will be used for tracking.  
As a result, the number of candidates finding the optimal parameters decreases to $42970$.
Fig. \ref{appendix_fig2} shows the mean accuracy in descending order. A large number of combinations yield high accuracy, implying that the proposed segmentation method is quite robust. 
Next, we find the parameters that appear most frequently within a specific range of sufficiently high accuracy.
The most appearing parameters were counted for $20000$ combinations (see the orange rectangle in Fig. \ref{appendix_fig2}).
The frequency of eight parameters is presented in Fig. \ref{appendix_fig3}, and the parameters which show the highest frequency are denoted by the yellow bars. 
We selected the values indicated by the yellow bars in Fig. \ref{appendix_fig3} as the optimal parameters. The segmentation presented in the paper was performed with these parameters. 
For space-time filtering, $\tau_{F}=0.25$, $K=100$, $\sigma=0.1$. For the local Otsu's method, $s=50$, $\delta=0.5$, and $\tau_{S}=0.25$, $K=10$, $\sigma=1$ for the SUBSURF method.

\end{appendices}

\clearpage
\section*{Author contributions}
\text{Seol Ah Park}, \text{Zuzana Kriv{\'a}}, and \text{Karol Mikula} designed the methods.
\text{Seol Ah Park} implemented, tested the methods, and performed data analysis.
\text{Seol Ah Park} and \text{Karol Mikula} wrote the manuscript, with feedback from \text{Georges Lutfalla} and \text{Mai Nguyen-Chi}.
\text{Tamara Sipka} and \text{Mai Nguyen-Chi} acquired and provided the data. 
\text{Georges Lutfalla} led the project.
\text{Mai Nguyen-Chi} and \text{Karol Mikula} co-led the project.
\bibliographystyle{unsrt}
\bibliography{reference}

\end{document}